\documentstyle[12pt]{article}
\def\be{\begin{equation}}
\def\te{\end{equation}}
\def\bea{\begin{eqnarray}}
\def\nn{\nonumber}
\def\tea{\end{eqnarray}}
\def\endignore{}
\def\beginignore#1\endignore{}
\def\a{\alpha}
\def\b{\beta}
\def\c{\raisebox{.4ex}{$\chi$}}
\def\d{\delta}
\def\e{\epsilon}
\def\f{\phi}
\def\g{\raisebox{.4ex}{$\gamma$}}
\def\h{\eta}
\def\i{\iota}
\def\j{\psi}
\def\k{\kappa}
\def\l{\lambda}
\def\m{\mu}
\def\n{\nu}
\def\o{\omega}
\def\p{\pi}
\def\q{\theta}
\def\r{\rho}
\def\s{\sigma}
\def\t{\tau}
\def\u{\upsilon}
\def\x{\xi}
\def\z{\zeta}
\def\D{\Delta}
\def\F{\Phi}
\def\G{\Gamma}
\def\J{\Psi}
\def\L{\Lambda}
\def\O{\Omega}
\def\P{\Pi}
\def\Q{\Theta}
\def\U{\Upsilon}
\def\X{\Xi}

\newskip\humongous \humongous=0pt plus 1000pt minus 1000pt
\def\caja{\mathsurround=0pt}
\def\eqalign#1{\,\vcenter{\openup2\jot \caja
        \ialign{\strut \hfil$\displaystyle{##}$&$
        \displaystyle{{}##}$\hfil\crcr#1\crcr}}\,}
\newif\ifdtup
\def\panorama{\global\dtuptrue \openup2\jot \caja
        \everycr{\noalign{\ifdtup \global\dtupfalse
        \vskip-\lineskiplimit \vskip\normallineskiplimit
        \else \penalty\interdisplaylinepenalty \fi}}}
\def\li#1{\panorama \tabskip=\humongous
        \halign to\displaywidth{\hfil$\displaystyle{##}$
        \tabskip=0pt&$\displaystyle{{}##}$\hfil
        \tabskip=\humongous&\llap{$##$}\tabskip=0pt
        \crcr#1\crcr}}
\def\eqalignnotwo#1{\panorama \tabskip=\humongous
        \halign to\displaywidth{\hfil$\displaystyle{##}$
        \tabskip=0pt&$\displaystyle{{}##}$
        \tabskip=0pt&$\displaystyle{{}##}$\hfil
        \tabskip=\humongous&\llap{$##$}\tabskip=0pt
        \crcr#1\crcr}}
 
\def\NP#1{Nucl. Phys.\ #1}
\def\PL#1{Phys. Lett.\ #1}
\def\PRL#1{Phys. Rev. Lett.\ #1}
\def\PRD#1{Phys. Rev. {\bf D} \ #1}
\def\PRA#1{Phys. Rev. {\bf A} \ #1}
\def\PTP#1{Prog. Theor. Phys.\ #1}
\def\AA#1{Astron. and Astrophys.\ #1}
\def\REF#1{$\sp{#1}$}
\def\pref#1{\hbox{$^\scriptsize{{\cite{#1}}}$}}
 
\def\vk{\bf k}
\def\vq{\vec q}
\def\bpm{\beta_{\pm}}
\def\bpl{\beta_+}
\def\bmn{\beta_-}
\def\ok{\omega_{\bf k}}
\def\a{\alpha}
\def\Alpha{\hat\alpha}
\def\Beta{\hat\beta}
\def\ha{{1\over 2}}
\def\dtr{\tilde{d}_R}
\def\ls{\stackrel{<}{\sim}}
\def\gs{\stackrel{>}{\sim}}
\def\nli{\nu_L^i}
\def\nri{\nu_R^i}

\newcommand{\Ubar}{\overline U}
\newcommand{\Dbar}{\overline D}
\newcommand{\Ebar}{\overline E}


\begin{document}
\begin{titlepage}
\begin{flushright}
{\large
UMD-PP-97-41\\
hep-ph/xxx\\
August 1996\\
}
\end{flushright}
\vskip 1cm
\begin{center}
{\Large\bf Baryon Non-Conservation in Unified Theories, } \\ \vskip .2cm
{\Large\bf in the Light of Supersymmetry and Superstrings
\footnote{Research supported by NSF grant No. PHY-9119745.  Invited
talk, presented at the Oak Ridge International Workshop on "Baryon
Instability", March, 1996.
Email: {\tt pati@umdhep.umd.edu}}}
\vskip 1cm
{\large
Jogesh C. Pati\\}
\vskip 0.5cm
{\large\sl Department of Physics\\
University of Maryland\\
College Park, MD~20742\\}
\end{center}
\vskip .5cm
\begin{abstract}
The first part of this talk presents the general complexion of baryon
and lepton number non-conservation that may arise in the context of
quark-lepton unification, and emphasizes the importance of searching
for both $(B-L)$-conserving proton decay modes--i.e. $p
\to \bar{\nu} K^+, \mu^+ K^o$, and $e^+ \pi^o$ etc.--as
well as $(B-L)$-violating transitions--i.e. $p \to e^- \pi^+ \pi^+,
n-\bar{n}$-oscillation and neutrinoless double beta decay.

The second part presents the status of grand unification with and
without supersymmetry and spells out the characteristic proton decay
modes, which if seen, will clearly show supersymmetry.  The {\it main
theme} of this talk, that follows next, pertains to two issues:  (i) the
need to remove the mismatch between MSSM and string-unifications; and
especially (ii) the need to resolve naturally the problem of rapid
proton decay, that generically arises in SUSY unification.
Seeking for a natural
solution to this second problem, it is noted that SUSY GUTS, including
SUSY SO(10) and $E_6$, can at best accommodate proton-stability by a
suitable choice of the Higgs-multiplets and discrete symmetries, but not
really explain it, because they do not possess the desired symmetries to
suppress both $d=4$ \underline{and} $d=5$ proton-decay operators.  By
contrast, following a recent work, I argue that a class of
string-solutions, possessing {\it three families}, does possess the desired
symmetries, which naturally safeguard proton-stability from all
potential dangers.  They also permit neutrinos to have desired light
masses.  This shows that, believing in supersymmetry, {\it superstring
is needed just to understand why the proton is so stable.}  Some
implications of the new symmetries, in particular the fact that they
still lead to observable rates for proton-decay in the same context in
which the mismatches between MSSM and string-unification are removed,
are noted.
\end{abstract}
\end{titlepage}

\section{Introduction}
\baselineskip=20pt
Non-conservation of baryon and lepton numbers, in particular proton
decaying into leptons, is one of the hallmarks of grand unification
\cite{1,2,3}.  In this context, other forms of baryon and lepton
non-conservation could be permitted as well -- such as neutrinoless
double beta decay and neutron-anti-neutron oscillation \cite{4}.
Unfortunately, experimental searches for any of these processes have not
yet produced a positive result \cite{5,6,7}.  Nevertheless, it is known
that some form of violation of baryon and/or lepton number must have
occurred in the early universe to account for the observed excess of
baryons over anti-baryons \cite{8,9}, which is in fact crucial to the
origin of life.  In addition, as discussed below, violation of at least
lepton number is strongly suggested by recent neutrino-oscillation
experiments \cite{10}, which indicate non-vanishing but light masses for
the neutrinos.  From a purely theoretical viewpoint, to be presented in
the following, it turns out that for a number of grand unified as well
as superstring-derived models, observation of proton decay, into modes
such as $\bar{\nu}_\mu K^+$ and/or $e^+ \pi^o$, should be within the
reach of current and forthcoming experimental facilities.

It thus seems most encouraging and timely that SuperKamiokande, with the
capability to improve the sensitivity of previous facilities - with
respect to searches for both proton decay and neutrino-oscillations --
by more than an order of magnitude, has just been turned on; and new
facilities like SNO and ICARUS, as well as those designed to search for
neutrino-less double beta decay and $n -\bar{n}$ oscillations with
improved sensitivity are expected to be available in the near future.

On the theoretical front, the original motivation for a unity of the
fundamental forces and that for questioning baryon and lepton-number
conservation laws in the context of such unification ideas \cite{1,2,3}
have remained unaltered.  But the perspective with regard to both issues
has changed significantly over the last two-and-a-half decades, owing to
the introduction of the ideas of supersymmetry \cite{11} and
superstrings \cite{12}.  In particular, supersymmetry, which seems to be
an essential ingredient for higher unification (see discussion in Sec.
5) poses the problem of rapid proton decay.  This is because, in accord
with the standard model gauge symmetry $SU(2)_L \times U(1)_Y \times
SU(3)^C$, a
supersymmetric theory in general permits, in contast to
non-supersymmetric ones, dimension 4 and dimension 5 operators which
violate baryon and lepton numbers \cite{13}.  Thus, unless these
operators are suppressed to the extent needed (see discussions later),
they pose the danger of unacceptably rapid proton decay.  It turns out
that obtaining a natural solution to this problem gets even harder if
one wishes to obtain at the same time non-vanishing but light masses for
the neutrinos ($\leq$ few eV)\cite{10}.

Bearing these issues in mind, I will first present here a brief summary
of the status of non-supersymmetric and supersymmetric grand
unificiation, and next a
current perspective on baryon- and lepton-number conservation laws in
the light of the ideas of supersymmetry and superstrings.  In this
latter part, following a recent work by me~\cite{14},
I will also present a natural solution to the problem of
rapid proton decay in the context of supersymmetry.   {\it It turns
out that the solution in question needs certain symmetries which can
not arise within conventional grand unification symmetries including
$E_6$, but they do arise within
superstring-derived three-family solutions.}  These
symmetries play an {\it essential role} in safeguarding
proton-stability from all potential dangers, to the extent desired, and
simultaneously permit neutrinos to have light masses of a nature that is
relevant to current experiments.  {\it This in turn provides a
strong motivation for symmetries of string-origin}.  The
extra symmetries in question lead to extra $Z^\prime -$ bosons, whose
currents bear the hallmark of string theories.  It turns out that there
is an interesting correlation between the masses of the $Z^\prime -$
bosons and observability of proton decay.

In section 2, I present the need for $B-L$ violation and that for
$SU(4)$-color.  In section 3, general complexion of $(B,L)$ violations and
the characteristic mass-scales associated with different processes are
listed, and in Section 4, the main ideas as regards physics beyond the
standard model are presented.  Section 5 provides the current status of
grand unification in the context of supersymmetry and raises
the issue of compatibility between MSSM and string-unification.
Certain attempts to
achieve this compatibility are presented.  In section 6, I present the
problem of $d=4$ and $d=5$ proton-decay operators and propose a solution
that naturally safeguards proton-stability from all potential dangers.
Some concluding remarks are presented in Section 7.

\section{$B-L$ Violation and $SU(4)$-Color}
As stated above, the observed excess of baryons over anti-baryons
implies that some form of violation of $B$ and/or $L$ must have occurred in
the early universe\cite{8,9}.  Such an excess could arise through $(B,L)$
violating processes which either conserve $B-L$, or violate it.  Kuzmin,
Rubakov and Shaposnikov pointed out, however, that any
excess generated through $(B-L)$-conserving processes at very early
moments of the universe (corresponding to temperatures $\gg 1 TeV$, i.e.
$t \ll 10^{-12} sec)$ is erased subsequently by purely electroweak
effects\cite{15}.  At the same time, generating baryon-excess through
electroweak effects alone does not seem to be adequate to account for
the observed baryon-asymmetry\cite{15}.  These considerations suggest
that baryogenesis must have its origin (at least in part) in processes
which violate $B-L$.

There is yet an {\it independent motivation} for violation of $B-L$
which stems from considerations of neutrino masses.  The reason is as
follows.  The simplest explanation for non-vanishing but light masses
for the neutrinos\cite{10} arises in the context of left-right symmetric
gauge theories\cite{16} and the so-called see-saw mechanism\cite{17}.
The minimal nonabelian version of a left-right symmetric gauge theory is
provided by the symmetry\cite{1}

\begin{equation}
{\cal G}_{224} \, = \, SU(2)_L \, \times \, SU(2)_R \, \times \, SU(4)^C
\end{equation}
which ensures (i) quantization of electric charge, (ii) quark-lepton
unification (through $SU(4)$-color), as well as (iii)
parity-conservation\cite{16}, at a basic level.  Any such theory
containing either $SU(2)_L \times SU(2)_R$ or $SU(4)^C$ necessarily
implies the existence of right-handed neutrinos $(\nu^i_R)$, accompanying
the left-hand ones $(\nu^i_L)$.  The see-saw mechanism\cite{17} assigns
{\it heavy Majorana masses} $(M^i_R \gg 1 TeV)$ to the
right-handed neutrinos, though not to the left-handed ones.  This
involves a breaking of left-right symmetry and thus parity
spontaneously at a high
scale\cite{16}.  Now, the Majorana masses for the $\nu^\prime_R s$, in
conjunction with the standard Dirac masses $m^i_D$, naturally yield very
light masses $(\ll m^i_D)$ for the known neutrinos:

\begin{equation}
m(\nu^i_L) \, \sim \, (m^i_D)^2/M^i_R \quad (i=e,\mu,\tau)
\end{equation}
Noting that the Dirac mass $m^i_D$ of the ith neutrino is expected to be
comparable to the mass of the ith up-quark (barring QCD renormalization
effects), it turns out that these masses for the ${\nu_L}^\prime s$ have
just the right pattern to be relevant to the neutrino-oscillation
experiments\cite{9,18} and to $\nu_\tau$ being hot dark matter, with

\begin{equation}
m(\nu^i_L) \, \sim \, (10^{-8})eV, \; 3 \times 10^{-3}eV, \; 1-10eV) \;
\; (i=e,\mu,\tau)
\end{equation}
\underline{if} $M^i_R \, \sim \, 10^{12}$ GeV, within a factor of
10 \cite{19}.  Heavy Majorana masses for ${\nu_R}^\prime s$, however,
needs spontaneous violation of lepton number L (with $\Delta B =0$) and
therefore of $B-L$ at a heavy intermediate scale.

We thus see that both baryogenesis and neutrino masses suggest the need
for microscopic violation of $B-L$.  One can argue that
spontaneous violation of $B-L$
becomes obligatory in theories in which it is gauged.  This is because,
in these theories, there is a massless spin-1 particle coupled to $B-L$.
Such a particle
would be inconsistent with the results of E\"{o}tvos-type
experiments \cite{20}, unless it acquires mass spontaneously.  Thereby, the
associated charge, in this case $B-L$, must be violated spontaneously
\cite{1}.
Now, the simplest symmetry that gauges $B-L$ is
$SU(4)$-color, which unifies quarks and leptons by using the idea that
lepton number is the fourth color \cite{1}.  {\it In short,
$(B-L)$-violation, is an integral feature of any theory containing
$SU(4)$-color.}

These considerations thus suggest that our very existence, requiring
baryogenesis and therefore violation of $B-L$, bears the footprints of
certain unification ideas - in particular that of quark-lepton
unification through a symmetry-like $SU(4)$-color.

\section{General Complexion of (B,L)-Violating Processes and Effective
Mass Scales}

The $(B,L)$ violating processes which conserve $B-L$ primarily involve only
proton decaying into an {\it anti-lepton} plus mesons:
e.g. $p \to e^+ \pi^o
, p \to \mu^+ K^o$, $p \to \bar{\nu} K^+, p \to \bar{\nu} \pi^+$ etc.
Once one permits violation of $B-L$, however, a
whole new set of processes can in general occur.  These include:  (i) a
nucleon decaying into a {\it lepton} plus mesons - i.e. $p \to e^-
\pi^+ \pi^+$ and $n \to e^- \pi^+$ etc., or proton decaying into a
lepton + lepton + antiplepton + mesons - e.g. $p \to e^- e^+ \nu \pi^+$,
(iii) Majorana masses for the neutrinos, (iv) neutrinoless double beta
decay and (v) $n-\bar{n}$ oscillation.

Now, Majorana masses for the right handed neutrinos, that are needed for
the see-saw mechanism \cite{17}, can arise
by introducing the pair of Higgs multiplets $\Delta_L$ and
$\Delta_R$ which transform as (3,1,10) and (1,3,10) of ${\cal G}_{224}$
or equivalently a single Higgs
multiplet 126 of SO(10) \cite{21}, which contains $\Delta_L$ and
$\bar{\Delta}_R$.
(An alternative choice of Higgs multiplets will be presented later.)
Parameters of the Higgs sector can be arranged such that the minimum of
the potential induces a large VeV $<\Delta_R> = v_R >> 1$ TeV,
while $<\Delta_L> \approx 0$ \cite{22}.  In the presence of the Yukawa
coupling $h_M({\nu_R}^T C^{-1} \nu_R \Delta_R + {\nu_L}^T C^{-1} \nu_L
\Delta_L) + hc$, such a VeV would induce a heavy Majorana mass for
$\nu_R$.  As mentioned before, this, in conjunction with the familiar
Dirac mass, yields a very light Majorana mass for $\nu_L$, (see eq.
(2)).  The VeV of ${\Delta}_R$ would also break ${\cal G}_{224}$ into the
standard model symmetry $SU(2)_L \, \times \, U(1)_Y
\, \times \, SU(3)^C$.  In this way, $<\Delta_R>$ breaks
lepton number $L$ and $(B-L)$, each by two units.
It also breaks parity and quark-lepton unification.

A specific set of diagrams which utilize $<\Delta_R> \neq 0$ and/or
Majorana masses for the neutrinos and thereby induce some of the (B-L)
violating processes mentioned above are shown in figs. 1, 2, and 3.  The
amplitudes for these processes would, of course, depend upon the
effective Yukawa, quartic and gauge couplings entering into the
respective vertices, as well as on the masses of the intermediate
particles - i.e. those of $\Delta_R$, the color-triplets and
color-octets $\xi_3$ and $\xi_8$, as well as $W_R$ and $\nu_R$ -
which enter into
figs. 1, 2, and 3.  Now, in minimal symmetry-breaking schemes of
left-right symmetric grand unification models, such as those based on
one- (or two-step)- breaking of SO(10), the masses of these intermediate
particles typically turn out to be either superheavy $\sim 10^{15}$ GeV,
or at least medium heavy $\sim 10^{12} GeV)$\cite{23}.  In this case,
it is easy to verify that the rates of all these $(B,L)$-violating
processes would be far too small to be observable.  For example, even if
effective Yukawa and quartic couplings are of order one, the amplitudes
for $qqq \to {\it l} + (q \bar{q})$ (Fig. 2a) and $qqq \to \bar{q} \bar{q}
\bar{q}$ (Fig. 3) would be of order $(1/M^5_{\rm eff}) \leq
(\frac{1}{10^{12}GeV})^5 \leq 10^{-60} GeV^{-5}$, where as one would
need these amplitudes to be greater than or of order of $10^{-30}
GeV^{-5}$ - i.e. $M_{\rm eff} \leq 10^6 GeV$ (say), for the corresponding
processes to have observable rates.  Roughly, a similar conclusion can be
drawn from a general operator analysis, based on construction of
effective invariant operators\cite{24} and dimensional estimate.  The
results of such estimates for the effective mass-scales that would be
necessary for the various $(B,L)$-violating processes to have observable
rates are shown in Table I.

It needs to be said that while minimal symmetry-breaking schemes
for SO(10) typically lead to
effective mass-scales which are considerably larger than those shown in
Table I and thus rates that are considerably smaller than what would be
observable, there exist viable models of the Higgs system, some
involving supersymmetry and thereby at least technically natural fine
tuning, where relatively low effective mass-scales of the type shown in
Table I and, therefore, observable rates for $n-\bar{n}$
oscillation \cite{25}
and/or proton decaying into lepton plus mesons are obtained \cite{26}.

Thus, I believe that, from a broader theoretical perspective, and also
because of the great significance of a positive result, if it should
show, experimental searches for both $(B-L)-$conserving, as well as the
$(B-L)-$violating processes shown in Table I, are strongly motivated.

\begin{table}
\centering
\begin{tabular}{|ll|l|c|}
\hline
& Processes & Selection Rules & Eff. Mass Scale  \\
\hline
I. & $p \to e^+ \pi^o, \bar{\nu} \pi^+$ & $\Delta B = \Delta L = -1$ & \\
& $n \to e^+ \pi^-$ & $\Delta (B-L) = 0$ & $\sim 10^{15} GeV$ \\
& $p \to \bar{\nu}_\mu K^+, \mu^+ K^o$ & & \\
\hline
II. & $p \to e^- \pi^+ \pi^+$, & $\Delta B = - \Delta L = -1$ & \\
& $n \to e^- \pi^+$ & $\Delta(B-L) = -2$ & $\sim 10^5 GeV$ \\
& $p \to e^- e^+ \nu \pi^+$, & $\Delta (B+L) = 0$ & \\
& $n \to e^- e^+ \nu$ & & \\
\hline
III. & $p \to e^- \nu \nu \pi^+ \pi^+$ & $\Delta B = - \frac{\Delta L}{3} =
-1$ & $\sim 10^5 GeV$ \\
& $n \to e^- \nu \nu \pi^+$ & $\Delta(B-L) = -4$ & \\
\hline
IV. & $p \to e^+ \bar{\nu} \bar{\nu}$ & $\Delta B = \frac{\Delta L}{3} =
-1$ & $\sim 10^{4.5} GeV$ \\
& & $\Delta(B-L) = +2$ & \\
\hline
V. & $n-\bar{n}$ & $\Delta B = 2, \Delta L = 0$ & $\sim 10^5 GeV$ \\
\hline
VI. & $nn \to ppe^- e^-$ & $\Delta B = 0, \Delta L = 2$ & $\sim 10^{4.5}
GeV$ \\
\hline
\end{tabular}
\caption{Effective Mass-Scales based on operator analysis and dimensional
estimates for the corresponding processes to have observable rates.}

\end{table}

To discuss these issues specifically in the context of grand
unification and supersymmetry, I will first present briefly in the
next section the motivations
for certain theoretical ideas involving physics beyond the standard
model, and then discuss grand unificaiton with supersymmetry in the
following section.

\section{Going Beyond the Standard Model}

The standard model of particle physics has brought a good deal of
synthesis in our understanding of the basic forces of nature
and has turned out to be brilliantly
successful in terms of its agreement with experiments.  Yet, as 
recognized for some time, it falls short as a
fundamental theory because it introduces some 19 parameters.  And it
does not explain
(i) the coexistence of the two kinds of matter:  quarks {\it
and} leptons; (ii) the coexistence of the electroweak {\it and} the
QCD forces with their hierarchical strengths $g_1 \ll g_2 \ll 
g_3$, as observed at low energies; (iii) quantization of electric charge;
(iv) family-replication;
(v) inter and intrafamily mass-hierarchies;
and (vi) the origin of diverse
mass scales that span over more than 27 orders of magnitude from
$M_{Planck}$ to $m_W$ to $m_e$ to $m_\nu$.
There are in addition the two most basic questions:  (vii) how does
gravity fit into the whole scheme, especially in the context of a good 
quantum theory?, and (viii) why is the cosmological constant so small or 
zero?  

These issues constitute at present some of the major puzzles of particle 
physics and provide motivations for contemplating new 
physics beyond the standard model which should shed light on them.  
The ideas which have been proposed over the last two-and-a-half decades
and which do show promise to 
resolve at least some of these puzzles include the following 
hypotheses:
 
(1)~~{\bf Grand Unification}: ~~ The hypothesis of grand 
unification \cite{1,2,3}, which proposes an underlying unity of the fundamental
particles and their forces, appears attractive
because it explains at once (i) the quantization of electric charge, (ii) the 
existence of quarks {\it and} leptons with $Q_e=-Q_p$, and (iii) the 
existence of the strong, the electromagnetic and the weak forces with
$g_3\gg g_2\gg g_1$ at low energies.
These are among the puzzles listed above and grand unification 
resolves all three.  By
itself, it does not address, however, the remaining 
puzzles listed above, including the issues of family replication and 
origin of mass-hierarchies.  

(2)~~{\bf Supersymmetry}:~~ This is the symmetry that
relates fermions to bosons \cite{11}.  As a local
symmetry, it is attractive because it implies the existence of gravity.
It has
the additional virtue that it 
helps maintain a large hierarchy in mass-ratios such as
$(m_{\phi}/M_U) \sim 10^{-14}$ and $(m_{\phi}/M_{p\ell}) \sim 10^{-17}$,
without
the need for fine tuning, provided, however, such ratios are put in by
hand. 
Thus it provides a technical resolution of the gauge hierarchy problem,
{\it but
by itself does not explain the origin of the large hierarchies}.

(3)~~{\bf Preonic Substructures with Supersymmetry}:
The idea \cite{27,28} that
quarks, leptons and Higgs bosons are composites of a {\it common} set of
constituents called ``preons,'' which possess supersymmetry, is still
unconventional.  On the negative side, the
preonic approach needs a few unproven, though not implausible,
dynamical assumptions as regards
the preferred direction of symmetry breaking and saturation of
the composite spectrum \cite{28}.  On the positive side, it has
the advantage that it is far more
economical in field-content and especially in parameters than
the conventional grand unification models, because it has no
elementary Higgs boson.  Second, and
most important, utilizing primarily the symmetries of the
theory (rather than detailed dynamics) and the
forbiddeness of SUSY-breaking, in the absence of gravity,
the preonic 
approach provides simple explanations for the desired protection of 
composite quark-lepton masses and at the same time for the origins of 
family-replication, inter-family mass-hierarchy and diverse mass scales
\cite{27}.
It also provides several testable predictions.
For this reason, I still keep an open mind about the
preonic approach.  To maintain a focus, however, I will assume
in the rest of this talk
that quarks, leptons and Higgs bosons are elementary.

(4)~~{\bf Superstrings}: ~~Last but not least, the idea of
superstrings \cite{12} proposes that the elementary entities are not truly
pointlike but are extended stringlike objects with sizes $\sim
(M_{Planck})^{-1} \sim 10^{-33}$ cm.
These theories (which may ultimately be just one) appear 
to be most promising in providing a unified
theory of all matter (spins 0, 1/2, 1, 3/2, 2, ...) and of
all the forces of nature including gravity.  Furthermore, by smoothing out 
singularities, they seem capable of yielding a well-behaved
quantum theory of gravity.  In principle, assuming that quarks, leptons 
and Higgs bosons are elementary, a suitable superstring theory
could also account for the origin of the three families and the Higgs 
bosons at the string unification scale, as well as explain all the
parameters of the standard model.  But in practice, this has not happened as 
yet.  
Some general stumbling blocks of string theories are associated with 
the problems of (i) a choice of the ground state (the vacuum)
from among the many solutions and (ii) understanding supersymmetry breaking. 

These provide in a nutshell motivations for physics beyond the standard
model, which, as it turns out, has strong implications for
non-conservation of baryon and lepton numbers.
The ideas listed above are, of course, not mutually
exclusive.  In fact the
superstring theories already comprise the idea of local 
supersymmetry and the central idea of grand unification.
In the following, I first
recall the status of conventional grand unification with supersymmetry
and then discuss the issue of $(B,L)$-nonconservation in the context of
these ideas.

\section{Grand Unification and Supersymmetry}
\subsection{The Need for SUSY}
It has been known for some time that
the dedicated proton decay searches at the IMB and Kamiokande
detectors\cite{5} and more recently the precision measurements of the
standard model coupling constants (in particular $sin^2 \theta_W)$ at
LEP\cite{29} put severe constraints on the idea of grand unification.
Owing to these constraints, the non-supersymmetric minimal $SU(5)$, and
for similar reasons, the one-step breaking minimal non-supersymmetric
$SO(10)$-model as well, are now excluded.\cite{30}  For example, minimal
non-SUSY $SU(5)$ predicts:  (i) $\Gamma(p \to e^+ \pi^o)^{-1} < 6 \times
10^{31} yr$ and (ii) $sin^2 \theta_W(m_Z)) \mid_{\bar{MS}} = .214 \pm
.004$, where as current experimental data show:  (i) $\Gamma {(p \to e^+
\pi^o)_{expt}}^{-1} > 6 \times 10^{32} yr$\cite{5} and (ii) $sin^2
\theta_W{(m_Z)_{expt}}^{LEP} = .2313 \pm .0003$\cite{29,30}.  The
disagreement with respect to $sin^2 \theta_W$ is reflected most clearly
by the fact that the three gauge couplings $(g_1,g_2$ and $g_3)$,
extrapolated from below, fail to meet by a fairly wide margin in the
context of minimal \underline{non-supersymmetric} $SU(5)$ (see Fig. 4).

But the situation changes dramatically if one assumes that the standard
model is replaced by the minimal supersymmetric standard model (MSSM),
above a threshold of about $1 TeV$.  In this case, the three gauge
couplings are found to meet\cite{31,32}, at least
approximately, provided $\alpha_3(m_Z)$ is not too low (see
figs. 4 and 5 and discussions below).  Their scale of meeting is given
by
\begin{equation}
M_X \, \approx \, 2 \, \times \, 10^{16} \quad ({\rm MSSM \; or \; SUSY
SU(5)})
\end{equation}
$M_X$ may be interpreted as the scale where a supersymmetric $GUT$ (like
minimal SUSY SU(5) or SO(10)) breaks spontaneously into the
supersymmetric standard model gauge symmetry $SU(2)_L \, \times \, U(1)
\, \times \, SU(3)^c$.  Both because a straightforward meeting of the
three gauge couplings (in accord with LEP data) is possible only
provided SUSY is assumed, and also because SUSY provides at least a
technical resolution of the gauge-hierarchy problem by preserving the
\underline{input} small ratio of $(m_W/M_X)$ in spite of quantum
corrections, SUSY has emerged as an essential ingredient for higher
unification.

With $M_X \sim 2 \times 10^{16} GeV$ and thus lepto-quark gauge boson
masses $\sim 10^{16} GeV$, as opposed to $2 \times 10^{14} GeV$ for
non-SUSY $SU(5)$, the \underline{dimension-6} gauge boson-mediated
proton-decay amplitude of order $g^2/{M_X}^2$ would lead to proton
lifetime of order $10^{37 \pm 1}$ years.  This is too long to allow observable
proton decay.  For SUSY grand unification, there are,however, new
contributions to proton-decay possibly from dimension $4$ and
necessarily from dimension $5$ operators \cite{13}.  These latter arise
due to exchange of color-triplet (anti-triplet) Higgsinos, which lie in
the $5(\bar{5})$ of SU(5), or in the 10 of SO(10).  Since they are
damped by just one power of the color-triplet Higgsino mass ${m_H}_C$,
these new contributions would lead to extra rapid proton decay, unless
${M_H}_C$ is sufficiently heavy.  For example, for SUSY SU(5) (with low
$tan \beta \leq 2.5)$, the experimental limit ($\geq 10^{32}$ yrs) on
$\Gamma(p \to {\bar{\nu}}_\mu K^+)^{-1}$ is met provided \cite{33}.
\begin{equation}
{m_H}_C \; \geq \; 2 \, \times \, 10^{16} GeV.
\end{equation}
It is interesting that the requirement of coupling-unification for SUSY
SU(5) puts an upper limit on ${m_H}_C$ of about $2.4 \times 1^{16}$ GeV
\cite{33}, which is barely compatible with the lower limit given above
(eq.(5)).

\subsection{SUSY GUT and Proton Decay Modes}  Leaving out for a moment
the issue of how to ensure naturally such
a large mass for the triplet, while its doublet partner is light $(\leq
1 TeV)$ (I will return to this issue in Sec. 6), if one takes the
attitude that the $d=4$ operators are forbidden by a discrete symmetry
or R-parity (see discussions later), and that
the parameters and/or the Higgs-spectrum and the couplings
for SUSY SU(5) or SUSY SO(10) can be arranged so that the triplet is
appropriately heavy \cite{34}, as noted above, proton decay would occur
primarily through the $d=5$ operators (rather than $d=6$),
with an observable rate $\sim
(10^{32}-10^{34} yrs)^{-1}$, which would be induced by the exchange of
color-triplet Higgsinos.  These bring a new complexion to proton
decay modes.

Owing to symmetry of the bosonic
components, the effective $d=5$ operators of the form QQQL/M in the
superpotential must involve at least two different families \cite{35}.
As a result, the non-vanishing operators relevant for proton decay,
which arise effectively through exchange of color triplets, are of the
form:
(a) $(\phi_{u,t,c}) \phi_{di} \phi_s {\phi_{\nu}}_{\mu}$ and (b) $\phi_u
\phi_{di} (\phi_{t,c}) \phi_{\mu}^-$.  These give rise to $d=5$
interactions which
are quadratic in both fermion and boson operatos.  They need to be
dressed by wino or gluino-exchange loops to yield effective four-fermion
proton-decay interactions of the form $qqql$.  Operators of class (a)
lead to decay modes such as (see figs. 6)
\begin{equation}
p \to {\bar{\nu}}_{\mu} K^+ , \quad n \to {\bar{\nu}}_{\mu} K^o
\end{equation}
and also (see fig. 7)
\begin{equation}
p \to {\bar{\nu}}_{\mu} \pi^+ \; and \; n \to {\bar{\nu}}_{\mu} \pi^o, etc.
\end{equation}
which arise primarily through wino-exchange loops.  Those of class (b)
give rise, through both wino and gluino-exchange, to charged antilepton
decay modes (see fig. 8):
\begin{equation}
p \to \mu^+ K^o , \quad p \to \mu^+ \pi^o, etc.
\end{equation}
Note that these do not include the canonical $p \to e^+ \pi^o$-mode,
which is induced, in $SUSY SU(5)$ or minimal $SUSY SO(10)$,
primarily by the exchange of heavy gauge bosons $(\sim
10^{16} GeV)$ and thus strongly suppressed.
Given the quark masses and the relevant mixing angles, it turns out that
one typically expects the rate of $p \to {\bar{\nu}}_{\mu} K^+$-mode to
be larger than that of both $p \to {\bar{\nu}}_{\mu} \pi^+$-mode, by about a
factor of 2-10, and of $p \to \mu^+ K^o$-mode by as much as two to
three orders of magnitude, and similarly for neutron-decay
\cite{35,36a,33}.  However, given the large top mass, which leads to
large $\tilde{t}-\tilde{u}$ mixing through renormalization group
corrections, it turns out that contribution from gluino-exchange can be
quite important, especially for large $\tan \beta \geq$ 40 (such large
$\tan \beta$ is permitted for SUSY SO(10) though not for SUSY SU(5)).  In
this case, $p \to \mu^+ K^o$ can compete favorably and perhaps even
dominate over the $p \to {\bar{\nu}}_{\mu} K^+$-mode \cite{36b}.  In
either case, we see that {\it one characteristic signal of SUSY GUT is
that strange particle decay modes -- i.e. $p \to {\bar{\nu}}_{\mu} K^+$
and/or $p \to \mu^+ K^o$ -- are at least prominent, and under some
circumstances dominant} \cite{35}.  There are regions in SUSY
parameter space, pertaining to the masses of the SUSY particles, the
mass of the color-triplets and $\tan \beta$, for which the non-strange
modes involving anti-leptons of the muon-family -- i.e. $p \to
{\bar{\nu}}_{\mu} \pi^+$ and possibly $p \to \mu^+ \pi^o$ -- can be
prominent or even dominant \cite{36a}, but those involving antileptons of
the first family -- in particular the $p \to e^+ \pi^o$ mode -- are
strongly suppressed.  These latter are, however, the dominant modes in
non-SUSY GUTS, like minimal non-$SUSY SU(5)$ and non-$SUSY SO(10)$.

{\it Thus observation of strange particle decay modes of the nucleon,
like $\bar{\nu} K^+$ or $\mu^+ K^o$,  as the
dominant or at least prominent modes, would clearly be a strong
signal in favor of supersymmetry.}  Furthermore, observation of certain
non-strange decay modes of the proton like ${\bar{\nu}}_{\mu} \pi^+$ or
$\mu^+ \pi^o$, as opposed to $e^+ \pi^o$, as prominent modes,
should also be a strong hint for the dominance of $d=5$
operators and thus for supersymmetry.

For the sake of completeness, it should be added, however, that there
exist \underline{SUSY} \underline{non-GUT} \underline{models},
for which even the $e^+
\pi^o$-mode can be prominent or even dominant.  That is the case, for
example, for the supersymmetric flipped $SU(5) \times U(1)'$-model
\cite{37a}, for which the mass of the relevant leptoquark gauge boson,
determined by the point of meeting of only $\alpha_3$ and $\alpha_2$
(though not ${\alpha_1}'$), can be much lower than $10^{16}$ GeV.  As
a result the gauge-boson mediated $d=6$-operator, which would lead to
$e^+ \pi^o$ as the dominant mode, can be the primary source of proton
decay and is likely to yield lifetimes of order $10^{32}-10^{34}$ yrs.
The $e^+ \pi^o$-mode can also arise as a prominent or dominant mode
through effective $d=4$ operators, which may be induced in
string-derived solutions through higher dimensional operators $(d > 5)$
utilizing VEVs of appropriate fields (see discussions in Sec. 6).  Thus,
observation of $e^+ \pi^o$ as a prominent or dominant decay mode of the
proton would certainly disfavor familiar SUSY GUTS, like $SUSY SU(5)$ or
$SUSY SO(10)$, but it would be perfectly compatible with the dominance
of {\it induced} $d=4$ operators, as mentioned above, or with the
flipped $SU(5)$ -model, and therefore with supersymmetry.

These points, as well as some related ones raised in the previous
section, show that {\it low-energy} studies of {\it selection-rules} for
$(B,L)$ non-conservation, pertaining to proton decay modes
as well as $n-{\bar{n}}$
oscillation, including a study of whether strange versus non-strange
decay modes of the nucleon are prominent, can
provide us with much information on possible new physics at very
short distances, spanning from $(100 TeV)^{-1}$ to
$(10^{17} GeV)^{-1}$.  I now return to the question of
unification of couplings in MSSM.

\subsection{MSSM-Unification and ${\bf \alpha_3}$}  Before entering into the
question of doublet-triplet spliting and a
natural suppression of the $d=4$ and $d=5$ proton decay operators, it
would be useful to probe into the accuracy with which the three
couplings meet and discuss the issue of a matching between MSSM and
string-unifications.  Minimal SUSY SU(5) predicts ${\rm sin^2
\theta_w}(m_Z)\mid_{\bar{MS}} = .2334 \pm .0036$, by using reasonable
range of masses for the SUSY particles, and more importantly a value
of $\alpha_3(m_Z) \, = \, .12 \, \pm \, .01$ as an input\cite{29,30,31}, where
the error bar in $\alpha_3 $ is more generous than that allowed by the
present world average data (see below).  Note that the predicted value
of $sin^2 \theta_W$ would agree with the observed one of $sin^2 \theta_W
(m_Z)_{expt} \, = \, .2313 \, \pm \, .0003$\cite{29}, only provided
$\alpha_3(m_Z)$, is {\it considerably higher than
.12}, which is what is reflected clearly by Fig. 5b (taken from Ref.
\cite{29}), if we demand a meeting of the three couplings.

This may be seen more succinctly by using the more accurately determined
value of $sin^2 \theta_W(m_Z) \, = \, .2313 \, \pm \, .0003$ as an input
and thereby getting $\alpha_3(m_Z)$ as an output\cite{29,32}.  If one
ignores possible corrections from GUT-threshold and Planck-scale
physics, it turns out that in this case one needs $\alpha_3(m_Z) \, >
\, .127$ to achieve coupling-unification within MSSM, assuming
$m_{\tilde{q}} \, \sim \, 1TeV$ and $m_{1/2} \, < \, 500 GeV$.
Such high values of $\alpha_3(m_Z) \, \geq \, .125$ (say) are, however,
incompatible with its world-average value \cite{37b},
\begin{equation}
\alpha_3(m_Z) \; = \; .117 \, \pm \, .005
\end{equation}
which is based on high as well as low-energy determinations of
$\alpha_3$.  The former is based on LEP-data and the latter on the
analysis involving $J/\psi, \Upsilon$, deep inelastic scattering and
lattice-calculations.

To summarize the situation with regard to coupling-unification, we see
that MSSM, which may be embedded in SUSY SU(5) or SUSY SO(10), fares far
better than non-SUSY GUT as regards achieving unification of the three
gauge couplings.  There does seem to be some discrepancy, however,
between the predicted
and the world-average values of $\alpha_3$, which, if genuine, would
imply that the three couplings do not quite meet at a point in the
context of MSSM.  The discrepancy may be resolved through large
corrections to the predicted $\alpha_3$ (as much as about $-.006$) which
may arise from GUT-threshold and Planck-scale physics.
Alternatively, the discrepancy may have its origin in {\it new
physics, beyond MSSM}, which may manifest at relatively low or
intermediate scale.  At this stage, not knowing precisely the
GUT-threshold and Planck-scale corrections, {\it it seems to me that one can
not discard MSSM, nor can one accept it necessarily as the whole truth,
representing the correct effective theory
below the GUT-scale.}  It turns out that a resolution of this issue gets
merged into a still bigger one pertaining to a matching between MSSM and
string-unifications, which in turn has implications on the precise
nature of (B,L)-nonconservation.  I therefore discuss next the issue of
this matching of unification from the two ends and the problem of low
$\alpha_3$.

\subsection{The Problem of Unification-Mismatch and Some Solutions}

Achieving a complete unity of the fundamental forces together with an
understanding of the origin of the three families and their
hierarchical masses is among the major challenges still confronting
particle physics.  Conventional grand unification, despite all the
merits noted in preceding sections, falls
short in this regard in that owing to the arbitrariness in the Higgs
sector, it does not even unify the Higgs exchange force, not to
mention gravity.  Superstring theory  is the only theory
we know that seems capable of removing these shortcomings.  It thus
seems imperative that the low energy data extrapolated to high
energies be compatible with string unification.  

It is, however, known \cite{38a,38b} that while the three gauge couplings,
extrapolated in the context of the minimal supersymmetric standard
model (MSSM) meet, at least aproximately\cite{32},
provided $\alpha_3(m_Z)$ is
not too low (as discussed above), their scale of meeting,
$M_X \approx 2 \times 10^{16}~GeV$, is nearly 20 times smaller than
the expected (one--loop level) string--unification scale \cite{39} of
$M_{\rm st} \simeq g_{\rm st} \times (5.2 \times 10^{17}~GeV) \simeq
3.6 \times 10^{17}~GeV$.  

Babu and I recently noted that very likely there is
a second mismatch concerning the value of the
unified gauge coupling $\alpha_X$ at $M_X$ \cite{40}.
Subject to the assumption of the MSSM spectrum, extrapolation
of the low energy data yields a rather low value of
$\alpha_X \approx 0.04$\cite{32},
for which perturbative physics should work well near $M_X$.  
On the other hand, it is known \cite{41}
that non--perturbative physics ought to be important for a
string theory near the string  scale, in order
that it may help choose the true vacuum and fix the moduli and the
dilaton VEVs.  The need to stabilize the dilaton in particular would suggest 
that the value of the unified
coupling at $M_{\rm st}$ in four dimensions should be considerably
larger than $0.04$ \cite{42}.  At the same
time, $\alpha_{\rm st}$ should not be too large, because, if
$\alpha_{\rm st} \gg 1$, the corresponding theory should be equivalent
by string duality \cite{43} to a certain
weakly coupled theory that would still suffer from the dilaton runaway
problem \cite{44}.  Furthermore, $\alpha_{\rm st}$ at $M_{\rm st}$ should not
probably be as large as even unity, or else, the one--loop string
unification relations for the gauge couplings \cite{39} would cease to
hold near $M_{\rm st}$ (e.g. in this case, the string threshold
corrections are expected to be too large) and the observed
(approximate) meeting of the three couplings would have to be viewed
as an accident.  {\it In balance, therefore, the preceding discussions
suggest that an
intermediate value of the string coupling $\alpha_{\rm st} \sim
.15-.25 (say) $ at $M_{\rm st}$ in four dimensions, which might be large
enough to stabilize the dilaton, but not so large as to disturb
significantly the coupling unification relations, is perhaps the more desired
value} \cite{40}.  In short, the desired unification of the
gauge couplings may well be {\it semi-perturbative}, rather than
perturbative, in character.  It is thus a
challenge to find a suitable variant or alternative to MSSM which
removes the mismatch not only with regard to the meeting point $M_X$, but
also with regard to the value of $\alpha_X$, as mentioned above.

A third relevant issue noted in the last section
is that the world average value of
$\alpha_3(m_Z) = 0.117 \pm 0.005$ \cite{37b} seems to
be low compared to
its value  that is needed for MSSM unification.  I note briefly a few
alternative suggestions which have
been proposed to address these issues, pertaining to
removing the mismatch between MSSM and string-unifications.

{\bf Matching Through String-Duality:}
One suggestion in this regard is due to
Witten \cite{45}.  Using the equivalence of the
strongly coupled heterotic $SO(32)$ and the $E_8 \times E_8$ superstring theories in 
$D=10$, respectively to the weakly coupled $D=10$ Type I and an $M$--theory, 
he observed that the
4-dimensional gauge coupling and $M_{\rm st}$ can both be
small, as suggested by MSSM extrapolation of the low energy data, without
making the Newton's constant unacceptably large.  While this observation opens
up a new perspective on string unification, its precise use to make
$\alpha_{\rm st} \approx 0.04$ at $M_{\rm st}$ would seem to run into
the dilaton runaway problem as in fact noted in Ref. \cite{45}.

{\bf Matching Through SUSY GUT:}  A second way in which the
mismatch between $M_X$ and $M_{\rm st}$
could be resolved is if superstrings yield an intact grand unification
symmetry like $SU(5)$ or $SO(10)$ with the right spectrum -- i.e.,
three chiral families and a suitable Higgs system including an adjoint
Higgs at $M_{\rm st}$ \cite{46}, and if this symmetry
would break spontaneously at $M_X \approx (1/20~ {\rm to}~ 1/50)
M_{\rm st}$ to the standard model symmetry.  However, as yet, there is
no realistic (or close-to realistic) string--derived GUT model \cite{46}.
Furthermore, for such solutions, there is the 
likely problem of doublet-triplet splitting and rapid
proton decay (see discussions later).

{\bf Matching Through Intermediate Scale Matter:}  A third
alternative is based on string--derived standard
model--like gauge groups.  It attributes the mismatch between $M_X$ and
$M_{\rm st}$ to the existence of new matter with intermediate
scale masses ($\sim 10^9-10^{13}$ GeV), which may emerge from
strings \cite{47}.  Such a resolution is
in principle possible, but it would rely on the delicate balance
between the shifts in the three couplings and on the existence of very
heavy new matter which in practice cannot be directly tested by experiments.
Also, within such alternatives, as well as those based on
non--standard hypercharge normalization \cite{48} and/or large
string--scale threshold effects \cite{49}, $\alpha_X$ typically remains
small ($\sim 0.04$), which is not compatible with the
need for a larger $\alpha_X$, as suggested above.

{\bf Matching Through ESSM -- An Example of Semi-Perturbative Unification:}
Babu and I recently noted \cite{40}
that all three issues raised above -- i.e. (i) understanding fermion
mass-hierarchy, (ii) finding a suitable alternative to MSSM which would
be compatible with string-unification, and (iii) accommodating low
$\alpha_3(m_Z)$ can have a {\it common resolution} through a
certain variant of the MSSM spectrum, which was proposed some time
ago \cite{27}.  The variant spectrum extends the
MSSM spectrum by adding to it two light
vector-like families $Q_{L,R} = (U,D, N, E)_{L,R}$ and
$Q'_{L,R}=(U',D',N',E')_{L,R}$, two Higgs singlets
($H_{\rm S}$ and $H_\lambda$)
and their SUSY partners, all as light as about 1 TeV and as heavy as
about 100 TeV \cite{50}.  We refer to this
variant as the Extended Supersymmetric Standard Model (ESSM).
The combined sets $(Q_L|{\overline{Q_R'}})$ and 
$({\overline{Q_R} }|Q_L'$) transform as ${\bf 16}$ and
${\bf \overline{16}}$ of $SO(10)$ respectively.
Barring addition of singlets, one can argue that ESSM is in fact
{\it the only extension} of the MSSM, containing complete families of
quarks and leptons, that is permitted by measurements of the oblique
electroweak parameters and $N_\nu$ on the one hand, and
renormalization group analysis on the other hand \cite{40}.  While the
derivation of such a spectrum in string theories, is not yet in hand,
it is worth noting
that the emergence of pairs of ${\bf 27} + {\bf \overline{27}}$ of $E_6$ or
${\bf 16} + {\bf \overline{16}}$ of $SO(10)$ in addition to chiral
multiplets is rather generic in string theories \cite{51,52}.

Babu and I performed a two-loop renormalization-group
analysis for the running of the three gauge couplings for
ESSM.  In this analysis, we included the contributions of the
Yukawa couplings of the two vector-like families with themselves and
with the three chiral families.  Such a pattern
of Yukawa coupling, which leads to a see-saw mass-matrix for the 3
chiral and two vector-like families, is motivated by
the inter-family mass-hierarchy \cite{27,53}.
All the relevant (unknown) Yukawa
couplings were assumed to have their fixed point values at the
electroweak scale, so that the analysis is essentially parameter-free,
except for the input gauge couplings and the variation in the
ESSM--spectrum.  Remarkably enough, we found that the three gauge couplings
$\alpha_{1,2,3}$ meet, even perfectly for many cases, for a fairly wide
variation in the ESSM spectrum.  A typical case of this meeting is shown
in Fig. 9.  The corresponding values
of $\alpha_X$, $M_X$ and $\alpha_3(m_Z)\mid_{\bar{MS}}$, with
vector-like quarks having masses $m_Q \approx 1.5-2 TeV$, are found to
be \cite{40}:
\begin{equation}
\alpha_X \, \approx \, .2-.25, \; M_X \, \approx \, 10^{17} GeV, \;
\alpha_3(m_Z)_{\bar{MS}} \, \approx \, .112-.118
\end{equation}
Raising $m_Q$ to $10-50~TeV$ would lower $\alpha_X$ to about .18 - .16, and
$M_X$ by about a factor of 2, leaving $\alpha_3$ in the range shown
above.

Thus we see that ESSM leads to coupling--unification, with an
intermediate value of $\alpha_X$, and a lower value of $\alpha_3(m_Z)$
than that needed for MSSM unification, just as desired.  The resulting
$M_X \sim 10^{17}~GeV$, though higher than the MSSM value, is
still lower than  the one--loop
string--unification scale of Ref.\cite{39}, which, for
$\alpha_X \approx 0.2$, yields $M_{\rm st} \approx 6 \times
10^{17}~GeV$.
Considering the proximity of $M_X \sim 10^{17}~GeV$ to the
expected string scale of $(5-8)\times 10^{17}~GeV$, however, it would
seem that contributions from
the infinite tower of heavy string-states, which have been neglected
in the running of $\alpha_i$'s, quantum gravity
and three and higher-loop effects \cite{A} may well play an
important role, especially for intermediate $\alpha_X \approx .2$,
in bridging the relatively small gap between $M_X$ and
$M_{\rm st}$.

As a general comment, with an intermediate unified coupling $(\alpha_X
\sim .2)$, the increased, though not overwhelming, importance of
three and higher loop-effects, compared to the case of MSSM, cannot of
course be avoided.  Yet, for such a case, the three gauge couplings are
still fairly weak ($< .15$, say) for most of the region of extrapolation
-- i.e. for $Q < 10^{15.5}$ GeV (say) (see Ref. \cite{40}).  As a result,
perturbation theory is still fairly reliable, all the way up to $M_X$,
and the benefits of calculability are not lost for this case, in
contrast to the case of a non-perturbative unification \cite{B}.
Thus, ESSM presents a good example of semi-perturbative unification,
that is viable, and also seems desirable, if one wishes to stabilize the
dilaton without losing the benefits of coupling-unification \cite{40}.
The main reason for devoting some discussion to these issues is that
intermediate $\alpha_X \sim .16 - .2$ turns out to be crucial to ensure
observable rate for proton decay in the same context in which rapid
proton decay is prevented.  This is what I discuss next.


\def\be{\begin{equation}}
\def\te{\end{equation}}
\def\bea{\begin{eqnarray}}
\def\nn{\nonumber}
\def\tea{\end{eqnarray}}

\def\a{\alpha}
\def\b{\beta}
\def\c{\raisebox{.4ex}{$\chi$}}
\def\d{\delta}
\def\e{\epsilon}
\def\f{\phi}
\def\g{\raisebox{.4ex}{$\gamma$}}
\def\h{\eta}
\def\i{\iota}
\def\j{\psi}
\def\k{\kappa}
\def\l{\lambda}
\def\m{\mu}
\def\n{\nu}
\def\o{\omega}
\def\p{\pi}
\def\q{\theta}
\def\r{\rho}
\def\s{\sigma}
\def\t{\tau}
\def\u{\upsilon}
\def\x{\xi}
\def\z{\zeta}
\def\D{\Delta}
\def\F{\Phi}
\def\G{\Gamma}
\def\J{\Psi}
\def\L{\Lambda}
\def\O{\Omega}
\def\P{\Pi}
\def\Q{\Theta}
\def\U{\Upsilon}
\def\X{\Xi}

\newskip\humongous \humongous=0pt plus 1000pt minus 1000pt
\def\caja{\mathsurround=0pt}
\def\eqalign#1{\,\vcenter{\openup2\jot \caja
        \ialign{\strut \hfil$\displaystyle{##}$&$
        \displaystyle{{}##}$\hfil\crcr#1\crcr}}\,}
\newif\ifdtup
\def\panorama{\global\dtuptrue \openup2\jot \caja
        \everycr{\noalign{\ifdtup \global\dtupfalse
        \vskip-\lineskiplimit \vskip\normallineskiplimit
        \else \penalty\interdisplaylinepenalty \fi}}}
\def\li#1{\panorama \tabskip=\humongous
        \halign to\displaywidth{\hfil$\displaystyle{##}$
        \tabskip=0pt&$\displaystyle{{}##}$\hfil
        \tabskip=\humongous&\llap{$##$}\tabskip=0pt
        \crcr#1\crcr}}
\def\eqalignnotwo#1{\panorama \tabskip=\humongous
        \halign to\displaywidth{\hfil$\displaystyle{##}$
        \tabskip=0pt&$\displaystyle{{}##}$
        \tabskip=0pt&$\displaystyle{{}##}$\hfil
        \tabskip=\humongous&\llap{$##$}\tabskip=0pt
        \crcr#1\crcr}}
 
\def\NP#1{Nucl. Phys.\ #1}
\def\PL#1{Phys. Lett.\ #1}
\def\PRL#1{Phys. Rev. Lett.\ #1}
\def\PRD#1{Phys. Rev. {\bf D} \ #1}
\def\PRA#1{Phys. Rev. {\bf A} \ #1}
\def\PTP#1{Prog. Theor. Phys.\ #1}
\def\AA#1{Astron. and Astrophys.\ #1}
\def\REF#1{$\sp{#1)}$}
\def\pref#1{\hbox{$^\scriptsize{{\cite{#1}}}$}}
 
\def\vk{\bf k}
\def\vq{\vec q}
\def\bpm{\beta_{\pm}}
\def\bpl{\beta_+}
\def\bmn{\beta_-}
\def\ok{\omega_{\bf k}}
\def\a{\alpha}
\def\Alpha{\hat\alpha}
\def\Beta{\hat\beta}
\def\ha{{1\over 2}}
\def\dtr{\tilde{d}_R}
\def\ls{\stackrel{<}{\sim}}
\def\gs{\stackrel{>}{\sim}}
\def\nli{\nu_L^i}
\def\nri{\nu_R^i}




\textheight=9.0in
\textwidth=6.2in
\topmargin=-.5in
\oddsidemargin=0.3in
\evensidemargin=0.in

\makeatletter                    
\makeatother                     

\section{The puzzle of proton$-$stability in Supersymmetry}

In this section, I first outline the problem of the unsafe $d=4$ as well
as the color-triplet mediated and/or gravity-linked $d=5$ proton decay
operators, and then, following a recent paper by me \cite{14},
present a solution which suppresses these unsafe operators naturally, so
as to ensure proton--stability, in accord with observation.  The
solution highlights the need for certain symmetries which
cannot arise in conventional grand unification, but which do arise in
string theories.



\subsection{The Problem and Attempted Solutions in SUSY GUTS}
As mentioned before, in accord with the standard model gauge symmetry
$SU(2)_L \times U(1)_Y \times SU(3)^C$, a supersymmetric theory in general
permits, in contrast to non-supersymmetric ones, dimension $4$ and dimension
$5$ operators which violate baryon and lepton numbers \cite{13}.
Using standard notations, the operators in question which may arise in the
superpotential are as follows:
\bea
W &=& [\eta_1 \Ubar \, \Dbar \, \Dbar  + \eta_2 Q L \Dbar + \eta_3 L L
\Ebar ] \nn\\
&+& [\l _1 QQQL + \l _2 \Ubar \, \Ubar \, \Dbar \, \Ebar  +
\l _3 LLH_2 H_2]/M.
\tea
Here, generation, $SU(2)_L$ and $SU(3)^C$ indices are suppressed.
$M$ denotes a characteristic mass scale.
The first two terms of $d=4$, jointly, as well as the $d=5$ terms of
strengths
$\lambda_1$ and $\lambda_2$, individually,
induce $\Delta(B-L) = 0$ proton decay with amplitudes
$\sim \h _1 \h _2/m_{\tilde
{q}}^{2}$ and $(\l _{1,2}/M)(\d )$ respectively, where $\d $
represents a loop-factor. Experimental limits on proton
lifetime turns out to impose the constraints:
$\h _1 \h _2 \leq 10^{-24}$ and $(\l _{1,2}/M)\leq 10^{-25}$ GeV$^{-1}$
\cite{55}.
Thus, even if $M \sim M_{string} \sim 10^{18}$ GeV, we must
have  $\l _{1,2} \leq 10^{-7}$,
so that proton lifetime will be in accord with experimental limits.

Renormalizable, supersymmetric standard-like and $SU(5)$ \cite{56}
models can be constructed so as to avoid,
{\it by choice}, the $d=4$ operators (i.e. the $\h _{1,2,3}$-terms) by
imposing a discrete or a multiplicative
$R$-parity symmetry: $R\equiv (-1)^{3(B-L)}$, or more
naturally, by gauging $B-L$, as in ${\cal G}_{224}
\equiv SU(2)_L\times SU(2)_R\times SU(4)^C$ or $SO(10)$.
Such resolutions,
however, do not in general suffice if we
permit higher dimensional operators and
intermediate scale VEVs of fields which violate $(B-L)$
and $R$-parity (see below). Besides, $B-L$ can not
provide any protection against the $d=5$ operators given
by the $\l_1$ and $\l_2$ - terms, which conserve $B-L$. These
operators are, however, expected to be present in any
theory linked with gravity, e.g. a superstring theory,
unless they are forbidden by some new symmetry.

As mentioned in Section 5, for SUSY grand unification models,
there is the additional problem that the exchange of
color-triplet Higgsinos which
occur as partners of electroweak doublets (as in ${\bf 5}
+{\bf \overline{5}}$
of $SU(5)$) induce $d=5$ proton-decay operators \cite{13}.
Thus, allowing for suppression of $\lambda_1$ and
$\lambda_2$ (by about $10^{-8}$) due to the smallness
of the relevant Yukawa couplings, the color-triplets still
need to be
superheavy ($\geq 10^{17}$ GeV) to ensure proton-stability
\cite{55},
while their doublet partners must be light ($\leq 1$ TeV).
This is the {\it generic problem of doublet-triplet splitting}
that faces all SUSY GUTS.  Basically, four types of solutions
to this problem have been proposed
in the context of SUSY grand unification.  They are as follows:
\newline (i) \underline{The Case of Extreme Fine Tuning}:  In this case,
utilizing cubic coupling in the superpotential of the form $W \, = \, A \,
\overline{5}_{H'} \, \cdot \, <24_H> \, \cdot \, 5_H \, + \, B \,
\overline{5}_{H'} \, \cdot \, <1_H> \, .5_H$, one can assign
masses to the triplets and the doublets in $\overline{5}$ and 5 of $SU(5)$
through the VEVs of both $24_H$ (which is traceless) and $1_H$.  By
arranging these two contributions to add for the triplets, but to almost
cancel for the doublets, {\it to one part in $10^{14}$}, one can
keep the doublets appropriately light, and the triplets
superheavy\cite{56}.  This case, however, needs extreme fine tuning.
\newline (ii) \underline{The Missing Partner Mechanism} \cite{57}:  In
this case, by introducing suitable large-size Higgs multiplets, such as
$50_H \, + \, \overline{50}_H \, + \, 75_H$, in addition to $5_H \, + \,
\overline{5}_H$ of $SU(5)$, and allowing couplings of the form $W \, =
\, C \, 5_H \, \cdot \, \overline{50}_H \, \cdot \, <75_H> \, + D \,
\overline{5}_H \, \cdot \, 50_H <75_H>$, one can give superheavy masses
to the triplets (anti-triplets) in $5(\overline{5})$ by pairing them with
anti-triplets (triplets) in $\overline{50}$(50).  But there do not exist
doublets in $50(\overline{50})$ to pair up with the doublets in
$5(\overline{5})$, which therefore can remain light, provided a direct
superheavy $5 \cdot \bar{5}$ mass-term is prevented.
\newline (iii) \underline{The Dimopoulos-Wilczek Mechanism} \cite{58}:
Utilizing the fact that the VEV of $45_H$ of $SO(10)$ does not have to
be traceless (unlike that of $24_H$ of $SU(5))$, one can give mass to
color-triplets and not to doublets in the $10$ of $SO(10)$, by arranging
the $VEV$ of $45_H$ to be proportional to diag $(x, x, x, o, o)$, and
introducing a coupling of the form $\lambda 10_{H1} \, \cdot \, 45_H \,
\cdot \, 10_{H2}$ in $W$.  Two $10's$ are needed owing to the
anti-symmetry of $45$.  Because of two $10's$, this coupling would leave
two pairs of electroweak doublets massless.  One must, however, make one
of these pairs superheavy, by introducing a term like $M \, 10_{H2}
\cdot 10_{H2}$ in
$W$, so as not to spoil the successful prediction of $sin^2\theta_W$ of
SUSY GUT.  In addition, one must
also ensure that only $10_{H1}$ but not $10_{H2}$ couple to the light
quarks and leptons, so as to prevent rapid proton decay.  All of these
can be achieved by imposing suitable
discrete symmeries.  There is, however, still some question as to
whether the triplets can be sufficiently heavy $(\geq 10^{17}) GeV)$
without conflicting with unification of the gauge couplings.  One can
avoid this issue altogether and ensure a strong suppression of
color-triplet mediated proton decay, provided one introduces additional
45's and 10 of SO(10) (see e.g. Babu and Barr, Ref. \cite{58}).
\newline (iv) \underline{The Case of Higgses as Pseudogoldstone Bosons
\cite{C}}.  A new line of approach, though not a complete model, has
been proposed recently, in which suitable choice of discrete symmetries
lead to accidental global symmetries of the Higgs-potential, which are
broken explicitly by the Yukawa couplings of the model.  The associated
pseudogoldstone bosons are identified with the Higgs doublets.  While
this idea has some nice features, because it proposes to use only
adjoint and fundamental representations for the Higgs scalars, the full
consistencey of this idea in the context of a complete model in
realizing electroweak-scale masses for the Higgs-doublets, and a
desirable pattern of masses for the fermions, remains to be shown.
Furthermore, in this case, one needs to make heavy use of discrete
symmetries to (a) ensure the accidental global symmetry of the
Higgs-potential, (b) obtain a desired pattern of masses for the
fermions, and (c) suppress undesirable flavor-changing neutral currents.
Thus, the prospect of a natural origin of this class of models (i.e. of
all the necessary discrete symmetries) from an underlying theory, like a
string theory, is far from clear.  Furthermore, the question of a
natural suppression of the $d=4$ -operators and of the gravity-linked
$d=5$ operators, of course, still remains open even in the context of
this class of models.

In summary, solutions to the problem of doublet-triplet splitting needing
either unnatural fine-tuning as in SUSY $SU(5)$ \cite{56},
or suitable {\it choice} of large number and/or large size
Higgs multiplets and/or choice of discrete symmetries
as in SUSY $SO(10)$ \cite{58}, missing-partner \cite{57} and
psuedogoldstone models \cite{C}, are technically
feasible.  {\it They, however, do not
seem to be compelling by any means because they  have been invented
for the sole purpose of suppressing proton-decay, without a
deeper reason for their origin.}  Furthermore, such solutions are
not easy to realize, and to date have not been realized, in
{\it string-derived} grand unified theories \cite{46}.

These considerations show that, in the
context of supersymmetry, the extraordinary stability of the
proton is a major puzzle.  {\it The question in fact arises:  Why does
the proton have a lifetime exceeding $10^{39}$ sec, rather than the
apparently natural value (for supersymmetry) of less than 1 sec?}  As
such, the known longevity of the proton deserves a natural
explanation.  Rather
than being merely accommodated, it ought to emerge as a
{\it compelling feature}, owing to symmetries of the underlying
theory, which should forbid, or adequately supress, the unsafe
$d=4$ as well as $d=5$ operators
in Eq. (11). As discussed below, the task of finding such symmetries
becomes even harder, if one wishes to assign
non-vanishing light masses ($\leq$ few eV) to
neutrinos.  Following Ref. 14, I will present in this section,
a class of solutions within supersymmetric theories, which {\it (a)
naturally} ensure proton-stability, to the extent desired, and
{\it (b)} simultaneously permit neutrinos to acquire light masses, 
of a nature that
is relevant to current experiments \cite{10,18}. These solutions need
\underline{either} $I_{3R}$ and $B-L$ as {\it separate} gauge
symmetries, as well as {\it one extra
abelian symmetry} that lies beyond even $E_6$ \cite{21};
\underline{or} the weak hypercharge
$Y$ ($=I_{3R} + (B-L)/2$) accompanied by {\it two extra symmetries}
beyond those of $E_6$. The interesting point is that while
the extra symmetries in question can not arise within conventional
grand unification models, including $E_6$, they do
arise within a class of
string-derived three generation solutions. This in turn
provides {\it a strong motivation}  for symmetries
of string-origin. The extra symmetries
lead to extra $Z'$-bosons, whose currents would bear
the hallmark of string theories. It turns out that there is an interesting
correlation between the masses of the $Z'$-bosons and observability
of proton decay.

\subsection{The need for symmetries beyond $SO(10)$ and $E_6$:} In what
follows, I assume that operators (with d $\geq 4$), scaled
by appropriate powers of Planck or string-scale mass, exist
in the effective superpotential of any theory which is
linked to gravity, like a superstring
theory \cite{12,59}, unless they are forbidden by the symmetries of the
effective theory.  For reasons discussed above, the class of
theories -- string-derived or not -- which contains $B-L$, as in
${\cal G}_{2311} \equiv 
SU(2)_L\times SU(3)^C\times U(1)_{I_{3R}}\times U(1)_{B-L}$, as a
symmetry, the $d=4$ operators in Eq.(11) are naturally forbidden.
They can {\it in general} appear however
through non-renormalizable operators if there exist VEVs of
fields which violate
$B-L$. This is where neutrino-masses
become relevant. As discussed in Sec. 2, the familiar see-saw mechanism
\cite{17} that provides the simplest reason for known neutrinos
$\nli $'s to be so light, assigns {\it heavy Majorana masses}
$M_R^i$ to the right-handed neutrinos $\nri $, which in turn need
spontaneous violation of $B-L$ at a heavy intermediate scale.

If $B-L$ is violated by the VEV of a field by two units, an
effective $R$-parity would
still survive \cite{60}, which would forbid the $d=4$ operators. That
is precisely the case for the multiplet
126 of $SO(10)$ or $(1,3,\overline{10})$ of ${\cal G}_{224}$,
which have commonly been used to give Majorana masses to
$\nu_R$'s. Recent works show, however, that $126$ and very likely
$(1,3,\overline{10})$, as well,  are hard
-- perhaps impossible -- to obtain in string 
theories \cite{61}. We, therefore, assume that this constraint holds.
It will become clear,
however, that as long as we demand safety from both $d=4$ {\it and}
$d=5$ operators, our conclusion as regards
the need for symmetries beyond $E_6$, would hold even if we give up
this assumption.

Without 126 of Higgs, $\nu_R$'s can still acquire heavy Majorana
masses utilizing product of VEVs of 
$s$neutrino-like fields $\widetilde{\overline{N_R}}$ and
$\widetilde{N'_L}$, which belong to 
$16_H$ and $\overline{16}_H$ respectively. (as in Ref. \cite{51},
see also \cite{52}.)
In this case, an effective operator of the form $ 16 \cdot 16 \cdot
\overline{16}_H.
\overline{16}_H/M$ in $W$, that is allowed by $SO(10)$,
would induce a Majorana mass  $(\overline{\nu}_R C^{-1}
\overline{\nu}_R^T)
(\langle\widetilde{N_L'}\rangle\langle\widetilde{N_L'}\rangle/M) 
+ hc\,$ of magnitude
$M_R\sim 10^{12.5}$ GeV, as desired, for
$\langle\widetilde{N'_L}\rangle \sim 10^{15.5}$ GeV and $M\sim 10^{18}$
GeV. \cite{62}
However, consistent with $SO(10)$ symmetry
and therefore its subgroups, one can have
an effective $d=5$ operator in the superpotential
$ 16^a \cdot 16^b \cdot 16^c \cdot 16_H/M $.
This would induce the terms $\Ubar _R \Dbar _R
\Dbar _R \langle\widetilde{\overline{N_R}}\rangle/M$ and  $QL\Dbar
\langle\widetilde{\overline{N_R}}\rangle/M$
in $W$ (see Eq.($11$)) with strengths
$\sim \langle\widetilde{\overline{N_R}}\rangle/M
\sim 10^{15.5}/10^{18} \sim 10^{-2.5}$, which would lead to unacceptably short
proton lifetime $\sim10^{-6}$ yrs.  We thus see that,
without having the $126$ or $(1, 3, \overline{10})$
 of Higgs,
{\it $B-L$ and therefore $SO(10)$ does not suffice
 to suppress even the $d=4$ - operators adequately
 while giving appropriate masses to neutrinos}. 
As mentioned before, $B-L$ does not of course prevent
the $d=5$, $\l_1$
and $\l_2$ - terms, regardless of the Higgs spectrum, because
these terms conserve $B-L$.

To cure the situation mentioned above, we need to utilize
symmetries beyond those of $SO(10)$. Consider first
the presence of at least one extra $U(1)$ beyond $SO(10)$ of the type
available in $E_6$, i.e. $E_6\rightarrow SO(10)\times U(1)_{\psi}$,
 under which
$27$ of $E_6$ branches into $(16_1 + 10_{-2} + 1_4)$, where $16$ contains
$(Q, L \mid \overline{U_R}, \overline{D_R}, 
\overline{E_R}, \overline{\nu_R})$, with $Q_{\psi}
= +1$; while $10$ contains the two Higgs doublets
$(H_1, H_2)^{(0, -2)}$ and a color-triplet and an anti-triplet
$(H^{(-2/3, -2)}_{\bf 3} + H^{'(2/3, -2)}_{3^{\ast}})$, where the
superscripts denote
$(B-L, Q_{\psi})$. Assume that the symmetry in the observable sector just
below the Planck scale is of the form:
\be
{\cal G}_{obs} = [{\cal G}_{fc} \subseteq SO(10)]
\times \hat{U}(1)_{\psi} \times [U(1)'s].
\te
It is instructive to first assume
that $\hat{U}(1)_{\psi} = U(1)_{\psi}$ of $E_6$ \cite{63}
and ignore all the other $U(1)$'s.
Ignoring the doublet-triplet splitting problem for a moment, 
we allow the 
flavor-color symmetry ${\cal G}_{fc}$ to be as big as $SO(10)$.
The properties of the operators in $W$ given in Eq.($7$), and
the fields
$\widetilde{\overline{N_R}}$, $(H_1, H_2)$ and the singlet $\chi\subset 27$,
under the charges $Y$, $I_{3R}$, $B-L$, $Q_{\psi}$ and
$Q_{T}\equiv Q_{\psi} - (B-L)$, are shown in Table $2$.
\begin{table}
\centering
\begin{tabular}{|c|c|c|c|c|c|}
\hline
Operators & $I_{3R}$ & $B-L$ & $Y$ & $Q_{\psi}$ & $Q_{T}$ \\
\hline
 & & & & & \\
$\Ubar \,\Dbar \,\Dbar$, $QL\Dbar$ & $1/2$ & -$1$ & $0$ & 3 & $4$ \\
$LL\overline{E}$ & $1/2$ & -$1$ & $0$ & $3$ & $4$ \\
$QQQL/M$ & $0$ & $0$ & $0$ & $4$ & $4$ \\
$\overline{U}\,\overline{U}\,\overline{D}\,\overline{E}/M$
& $0$ & $0$ & $0$ & $4$ & $4$ \\
$LLH_2H_2/M$ & $1$ & -$2$ & $0$ & -$2$ & $0$  \\
 & & & & & \\ \hline
 & & & & & \\
$\widetilde{\overline{N_R}}$ & -$1/2$ & $1$ & $0$ & $1$ & $0$ \\
$(H_1, H_2)$ & (-$1/2, 1/2)$ & $0$ & (-$1/2, 1/2)$ & -$2$ & -$2$ \\
$\chi$ & $0$ & $0$ & $0$ & $4$ & $4$ \\
 & & & & & \\ \hline
\end{tabular}
\caption{}
\end{table}
We see that the $d=4$ operators ($\eta_i$ -terms) carry nonvanishing
$B-L$, $Q_{\psi}$ and $Q_T$, and are thus forbidden by each of these
symmetries.  Furthermore, note that when
$\widetilde{\overline{N_R}}\subset 16$ and
$\widetilde{N'_L}\subset\overline{16}$ acquire VEV, the charges $I_{3R}$,
$B-L$ as well as $Q_{\psi}$ are broken, {\it but $Y$ and $Q_T$ are preserved}.
Now $Q_T$ would be violated by the VEVs of $(H_1, H_2) \sim
200$ GeV and of the singlets $\chi^{(27)}$ and
$\overline{\chi}^{(\overline{27})}$. 
Assume that $\chi$ and $\overline{\chi}$
acquire VEVs $\sim 1$ TeV through a radiative mechanism,
utilizing Yukawa interactions, analogous to
$(H_1, H_2)$. The $d=4$ operators can be induced
through nonrenormalizable terms of the type 
$16 \cdot 16 \cdot 16 \cdot [\langle\widetilde{
\overline{N_R}}\subset 16\rangle/M]. 
[\langle10\rangle\langle10\rangle/M^2$ or $\langle\overline{\chi}\subset
\overline{27}\rangle/M]$, where the effective couplings 
respect $SO(10)$ and $U(1)_{\psi}$.
Thus we get $\eta_i \leq (10^{15.5}/
10^{18})(1$ TeV$/10^{18}$ GeV)$\sim 10^{-18}$, which is below the
limit of $\eta_1 \eta_2 \leq 10^{-24}$. Thus, $B-L$
{\it and} $Q_{\psi}$, arising within $E_6$, suffice to control the
$d=4$ operators adequately, while permitting neutrinos to have
desired masses. 

Next consider the
$LLH_2H_2$-term. While it violates $I_{3R}$, $B-L$ and $Q_{\psi}$, it is
the only term that is allowed by $Q_T$. Such a term can arise through an 
effective interaction of the form $16 \cdot 16 \cdot (H_2\subset10)^2
\cdot \langle\widetilde{\overline{N_R}}\subset 16\rangle^2/M^3$, and 
thus with a strength
$\sim 10^{-5} \cdot (10^{18}$GeV)$^{-1}$, which is far below the limits
obtained from $\nu$-less double
$\beta$-decay.

Although the two $d=5$ operators $QQQL/M$ and
$\overline{U}\,\overline{U}\,\overline{D}\,\overline{E}/M$
are forbidden by $Q_{\psi}$ and $Q_T$, the problem of
these two operators still arises as follows. Even for a
broken $E_6$-theory, possessing $U(1)_{\psi}$-symmetry, the
color-triplets $H_3$ and $H'_{3^\ast}$ of 27 still exist in the
spectrum. They are in fact needed to cancel
the anomalies in $U(1)_{\psi}^3$ and $SU(3)^2 \times U(1)_{\psi}$
etc. They acquire masses of the form $M_3 H_3 H'_{3^\ast}
+ hc$ through the VEV of singlet $\langle \chi \rangle$ which
breaks $Q_{\psi}$ and $Q_T$ by
four units. With such a mass term, the exchange of these
triplets would induce $d=5$ proton-decay operators, just as it does
for SUSY $SU(5)$ and $SO(10)$. We are then back to facing either
the problem of doublet-triplet splitting (i.e. why $M_3 \geq 10^{17}$ GeV)
or that of rapid proton-decay (for $M_3 \sim 1$ TeV). {\it In this sense,
while the
$E_6$-framework, with $U(1)_{\psi}$, can adequately control the
$d=4$ operators and give appropriate masses to the neutrinos
(which $SO(10)$ cannot), it does not suffice to
control the $d=5$ operators, owing to the presence of
color-triplets.}  As we discuss below, this is where string-derived
solutions help in preserving the benefits of a $Q_{\psi}$-like charge,
while naturally eliminating the dangerous color-triplets.

\subsection{Doublet-Triplet Splitting In String Theories:  A
Preference For Non-GUT Symmetries:}
While the problem of
doublet-triplet splitting does not have a compelling solution
within SUSY GUTS and has not been resolved within string-derived
GUTS \cite{46},  it can be solved quite simply
within string-derived
standard-like \cite{64,65} or the ${\cal G}_{224}$-models \cite{51},
{\it because in these models, the electroweak doublets
are naturally decoupled from the
color-triplets after string-compactification}. As a result,
invariably, the same set of boundary conditions
(analogous to ``Wilson lines'') which break $SO(10)$ into a standard-like
gauge symmetry such as ${\cal G}_{2311}$, either project out,
by GSO projections, all
 color-triplets $H_3$ and $H_{3^\ast}^{'}$ from the
``massless''- spectrum \cite{65}, or yield some color-triplets
with extra $U(1)$ - charges which make them harmless
\cite{64}, because they can not have Yukawa couplings with quarks and leptons.
In these models, the doublet triplet splitting problem is
thus solved from the start, because the {\it dangerous} color -
triplets simply do not appear in the massless spectrum \cite{66}.

At the same time, owing to constraints of string theories,
the coupling unification relations hold \cite{39} for the
string-derived standard-like or ${\cal G}_{224}$-models, just as
they do for GUT-models.
Furthermore, close to realistic models have been derived
from string theories only in the context of such standard-like
\cite{64,65}, flipped $SU(5) \times U(1)$ \cite{37a}
and ${\cal G}_{224}$ models \cite{51}, but not yet for
string-derived GUTS \cite{46}.
For these reasons, we will consider string-derived non-GUT models,
as opposed to string-GUT-models, {\it as the prototype
of a future realistic string model}, and use them as a {\it guide}
to ensure {\it (a)} proton -
stability and {\it (b)} light neutrino masses \cite{67a}.

Now, if we wish to preserve the benefits of the charge $Q_{\psi}$
(noted before), and still eliminate the color-triplets
as mentioned above, there would appear to be a problem, because,
without the color-triplets,
the incomplete subset consisting of 
$\{16_1 + (2,2,1)_{-2} + 1_4\}\subset 27$ of $E_6$
would lead to anomalies in $U(1)_{\psi}^3$,
$SU(3)^2\times U(1)_{\psi}$ etc. This is where symmetries of
string-origin come to the rescue.

\subsection{Resolving the Puzzle of Proton-longevity through
string-symmetries}
The problem of anomalies (noted above) is cured within string
theories in a variety of ways. For instance,
new states beyond those in the $E_6$-spectrum invariably appear
in the string-massless sector which
contribute toward the cancellation of anomalies, and only certain
{\it combinations} of generators become {\it anomaly-free}.
To proceed further, we need to focus on some specific
solutions. For this purpose, we choose to explore
the class of string-derived three generation models, obtained
in Refs.\cite{64} and \cite{65}, which is as close
to being realistic as any other such model that exists in the
literature (see e.g. Refs. \cite{51} and \cite{37a}).
In particular, they seem capable of generating
qualitatively the right texture for fermion mass-matrices and CKM mixings.
We stress, however, that the essential feature of our
solution \cite{14}, relying primarily on the existence of extra
symmetries analogous to $U(1)_{\psi}$,
is likely to emerge
in a much larger class of string-derived solutions.

After the application of all GSO projections, the gauge symmetry
of the models developed in these references, at the string
scale, is given by:
\be
{\cal G}_{st} = [SU(2)_L \times SU(3)^C \times U(1)_{I_{3R}}
\times U(1)_{B-L}]
\times [G_M = \prod_{i=1}^{6} U(1)_{i}] \times G_H.
\te
Here, $U(1)_i$ denote six horizontal-symmetry charges which
act non-trivially on the three families and distinguish between
them.  In the models
of Refs. \cite{64}, \cite{65}, $G_H = SU(5)_H \times SU(3)_H \times
U(1)_H^2$. There exists ``hidden'' matter which couples
to $G_H$ and also to $U(1)_i$.

A partial list of the massless states for the solution
derived in Ref. \cite{64}, together with the associated $U(1)_i$-charges,
is given in Table 3.  The table reveals the following features:

{\it (i)} There are three families
of quarks and leptons (1, 2 and 3), each with $16$ components,
including $\overline{\nu_R}$. Their quantum numbers under the symmetries
belonging to $SO(10)$
are standard and are thus not shown. Note that the
$U(1)_i$ charges differ from one family to the other. There are
also three families of hidden sector multiplets $V_i$, $\overline{V}_i$,
$T_i$ and $\overline{T}_i$ which possess $U(1)_i$-charges.

{\it (ii)}
The charge $Q_1$ has the same value ($\frac{1}{2}$) for all
sixteen members of family $1$, similarly $Q_2$ and $Q_3$ for
families 2 and 3 respectively. In fact, barring a normalization
difference of a factor of 2, {\it the sum $Q_+ \equiv Q_1 + Q_2 +
Q_3$ acts on the three families and on the three Higgs doublets
$\overline{h_1}$, $\overline{h_2}$ and $\overline{h_3}$ in the
same way as the $Q_{\psi}$ of $E_6$ introduced before}. The
analogy, however, stops there, because the solution has
additional Higgs doublets (see table) and also because
there is only one pair of color triplets ($D_{45}, \overline{D}_{45}$)
instead of three. Furthermore, the pair
($D_{45}, \overline{D}_{45}$) is {\it vector-like} with
opposite $Q_+$-charges, while
($H_3, H'_{3^\ast}$), belonging
to $27$ of $E_6$, have the same $Q_{\psi}$-charge.
In fact the pair ($D_{45}, \overline{D}_{45}$) can have
an invariant mass conserving all $Q_i$-charges, but ($H_3, H'_{3^\ast}$)
can not.

{\it (iii)} It is easy to see that owing to different
$U(1)_i$-charges, the color-triplets
$D_{45}$ and  $\overline{D}_{45}$ (in contrast to $H_3$ and
 $H'_{3^\ast}$) can not have allowed Yukawa couplings to
 ($qq$) and ($ql$) - pairs. Thus, as mentioned before, they
 can not mediate proton decay.

{\it (iv)} Note that the solution yields altogether
four pairs of electroweak Higgs doublets: ($h_1, h_2, h_3,
h_{45}$) and ($\overline{h_1}, \overline{h_2}, \overline{h_3},
\overline{h_{45}}$). It has been shown \cite{64} that only one
pair -- i.e. $\overline{h_1}$ or $\overline{h_2}$ and
$h_{45}$ -- remains light, while the others acquire superheavy
or intermediate scale masses.  Owing to differing
$U(1)_i$-charges, the three families have Yukawa couplings
with three distinct Higgs doublets. Since only one pair
($\overline{h_1}$ and $h_{45}$) remains light and acquires
VEV, it turns out that families 1,2 and 3 get identified with the
$\tau$, $\mu$ and $e$-families respectively \cite{64}. The mass-heirarchy
and CKM mixings arise through higher dimensional
operators, by utilizing VEVs of appropriate fields and
hidden-sector condensates.

Including contributions from the entire massless spectrum,
one obtains: $Tr U_1 = Tr U_2 = Tr U_3 = 24$ and $Tr U_4 =
Tr U_5 = Tr U_6 = -12$. Thus, all six $U(1)_i$'s are
anomalous. They give rise to five anomaly-free combinations
and one anomalous one:
\bea
U'_1 &=& U_1 - U_2 \,,~~ \, U'_2 = U_4 - U_5 \,,~~ \,
 U'_3 = U_4 + U_5 - 2U_6 \,,\nn \\
\hat{U}_{\psi} &=& U_1 + U_2 - 2U_3 \,,\nn \\ 
\hat{U}_{\chi} &=& (U_1+U_2+U_3) + 2(U_4+U_5+U_6) \,,\nn \\
U_A &=& 2(U_1+U_2+U_3) - (U_4+U_5+U_6).
\tea
One obtains $Tr Q_A = 180$ . The anomalous $U_A$ is broken by
the Dine-Seiberg-Witten (DSW) mechanism, in which the
anomalous D-term generated by the VEV of the dilaton field
is cancelled by the VEVs of some massless fields which
break $U_A$, so that supersymmetry is preserved \cite{67}. The solutions
(i.e. the choice of fields with non-vanishing VEVs) to
the corresponding F and D - flat conditions are, however, not
unique. A few alternative possibilities have been
considered in Ref. \cite{64} (see also Refs. \cite{51} and \cite{37a}
for analogous considerations).
Following our discussions in the previous subsection
as regards non-availibility
of $126$ of $SO(10)$ or ($1,3,\overline{10}$) of ${\cal G}_{224}$, I
assume, for the sake of simplicity in estimating strengths
of relevant operators, that $B-L$ is violated
spontaneously at a scale $\sim 10^{15}$-$10^{16}$
GeV by one unit (rather than two)
through the VEVs of {\it elementary} $s$neutrino-like
fields $\widetilde{\overline{N_R}}
\subset 16_H$ and $\widetilde{N_L'} \subset \overline{16_H}$
(as in Ref. \cite{51}).  Replacing VEVs of these elementary
fields by those of products of fields including condensates,
as in Ref. \cite{64},
would only lead to further suppression of the relevant unsafe
higher dimensional operators and go towards strengthening our
argument \cite{14} as
regards certain symmetries being sufficient in preventing
rapid proton-decay \cite{68}.


\noindent {\bf A Longlived Proton:}
We now reexamine the problem of proton-decay and neutrino-masses
by assuming that in addition to $I_{3R}$ and
$B-L$, or just $Y$, either $\hat{Q}_{\psi} \equiv Q_1 + Q_2 - 2 Q_3$,
or $\hat{Q}_{\chi} \equiv Q_1+ Q_2 + Q_3 + 2( Q_4 + Q_5 + Q_6 ) $
(see Eq. (10)), or both emerge as good symmetries near the string scale,
which are broken by the VEVs of (i) $s$neutrino-like fields $\sim 10^{15}$
- $10^{16}$ GeV, (ii) electroweak doublets and
singlets (denoted by $\phi$'s)
$\sim 1 TeV$, and (iii) hidden-sector condensates.
To ensure proton-stability, we need to assume
that the hidden-sector condensate-scale is $\leq 10^{-2.5} M_{st}$.
With the gauge coupling $\alpha_X$, at the unification-scale
$M_X$, having nearly the MSSM value of $.04 - .06$, or even
an intermediate
value $\approx .16 - .2$ (say), as suggested in Ref. \cite{40},
this seems to be a safe assumption for most string models
(see discussions later).
The roles of the symmetries $Y$, $B-L$, $\hat{Q}_{\psi}$, $\hat{Q}_{\chi}$
and ($\hat{Q}_{\chi} + \hat{Q}_{\psi}$)
in allowing or forbidding the relevant $(B,L)$ - violating
operators, including the higher dimensional ones, which allow violations
of these symmetries through appropriate VEVs,
are shown in Table 4.  Based on the entries in this table,
the following points are worth noting:


\noindent {\it (i)} \underline{Inadequacy of the Pairs ($Y$, $B-L$);
($Y$, $\hat{Q}_{\psi}$); ($Y$, $\hat{Q}_{\chi}$) and ($B-L$,
$\hat{Q}_{\chi}$)}: Table 4 \\
shows that no single charge nor the pairs ($Y, B-L$), ($Y$, $\hat{Q}_{\psi}$),
($Y$, $\hat{Q}_\chi$) and ($B-L$, $\hat{Q}_\chi$) give adequate
protection against all the unsafe operators.
Let us next consider other pairs of charges.

\noindent {\it (ii)} \underline{Adequate Protection Through the
Pair ($B-L$ and $\hat{Q}_{\psi}$) or the Pair
($\hat{Q}_{\chi}$ and $\hat{Q}_{\psi}$)}: \\
Using Table 4, we observe that
the pair ($B-L$ and $\hat{Q}_{\psi}$), as well as
the pair ($\hat{Q}_{\chi}$ and $\hat{Q}_{\psi}$), forbid
all unsafe operators, including
those which may arise from higher dimensional ones, with or without
hidden-sector condensates. In fact,
{\it members of the pairs mentioned above complement each other
in the sense that when one member of a pair allows an unsafe
operator, the other member of the same pair forbids it, and
vice versa} -- a remarkable team effort.
Note that the strengths
of the d=4 and d=5 operators are controlled by the VEVs 
$ \, < \overline{h}_1/M >^2$, 
$< \Phi/M >^n$ and 
$< T_i \overline{T}_j/M^2 >^2$,
which give more than necessary suppression, even if the condensate-scale
is as large as about $10^{15}$ GeV (see estimates below).

\noindent {\it (iii)} \underline{$\hat{Q}_{\psi}$ removes Potential
Danger From Triplets in The Heavy Tower As Well}:
Color \\
triplets in the heavy infinite tower of
states with masses $M\sim M_{st} \sim 10^{18}$ GeV in general
pose a {\it potential danger} for all string theories, including
those for which they are projected out
from the massless sector \cite{64}. The exchange of these heavy
triplets, if allowed, would induce $d=5$ proton-decay operators
with strengths $\sim \kappa/M$, where $\kappa$ is given by
the product of two Yukawa couplings. Unless the Yukawa couplings are
appropriately suppressed \cite{69} so as to yield $\kappa \leq 10^{-7}$
\cite{55}, these operators would be unsafe. {\it Note, however,
that string-derived solutions possessing
symmetries like $\hat{Q}_{\psi}$ are free from this type
of danger}. This is because, if $\hat{Q}_{\psi}$ emerges as a good
symmetry near the string-scale, then the spectrum, the masses and the
interactions of the color-triplets in the heavy tower would respect
$\hat{Q}_{\psi}$. As a result, the exchange of such states can not
induce $d=5$ proton-decay operators, which violate $\hat{Q}_{\psi}$
(see Table 4).

In fact, for such solutions, the color-triplets in
the heavy tower can appear only as {\it vector-like pairs},
with opposite $\hat{Q}_{\psi}$-charges (like those in
$10$ and $\overline{10}$ of $SO(10)$, belonging to
$27$ and $\overline{27}$ of $E_6$ respectively), so that they can
acquire invariant masses of the type $M\{(H_3 \overline{H_3} +
H'_{3^\ast}\overline{H'_{3^\ast}}) + hc\}$, which conserve
$\hat{Q}_{\psi}$. Such mass-terms cannot induce
proton decay.

Thus we see that a symmetry like $\hat{Q}_{\psi}$
plays an essential role in safegaurding proton-stability
from all angles \cite{14}.
Since $\hat{Q}_{\psi}$
distinguishes between the three families \cite{70}, it cannot, however,
arise within single - family grand unification
symmetries, including $E_6$. 
{\it But it does arise within string-derived
three-family solutions (as in Ref. \cite{64}), which
at once know the existence of all three families}. In this
sense, string theory plays a vital role in explaining
naturally why the proton is so extraordinarily stable, in spite
of supersymmetry, and why the neutrinos are so light.

\noindent{\bf$Z'$-mass and proton decay rate}:  If symmetries
like $\hat{Q}_{\psi}$ and possibly
$\hat{Q}_\chi$, in addition to $I_{3R}$ and $B-L$, emerge as
good symmetries near the string scale, and break
spontaneously so that only electric
charge is conserved, there must exist at least one
extra $Z'$-boson (possibly more),
in addition to the (almost) standard $Z$ and a superheavy $Z_H$
(that acquires a mass through $s$neutrino--VEV)
\cite{71}. The  extra $Z'$ boson(s) will be associated with
symmetries like $\hat{Q}_T \equiv 2\hat{Q}_{\psi} - (B-L)$ and
$\hat{Q}_\chi + \hat{Q}_{\psi}$, in addition to $Y$, that
survive after $s$neutrino acquires a VEV. The $Z'$ bosons can
acquire masses through the VEVs of electroweak doublets and singlets
($\phi$'s), as well as through the hidden-sector condensates like
$\langle\overline{T}_i T_j\rangle$, all of which break $\hat{Q}_T$
and $\hat{Q}_\chi + \hat{Q}_{\psi}$ (see Table 3).
As mentioned before, we
expect the singlet $\phi$'s to acquire VEVs, at least radiatively
(like the electroweak doublets), by utilizing their Yukawa couplings
with the doublets, which at the string-scale is comparable to
the top-Yukawa coupling. Since the $\phi$'s do not
have electroweak gauge couplings, however, we would expect that
their radiatively-generated VEV, collectively denoted by $v_0$, to
be somewhat higher than those of the doublets ($v_{EW}\sim 200$ GeV)
-{\it i.e.}, quite plausibly, $v_0 \sim 1$ TeV.
Thus, in the absence of hidden-sector condensates,
we would expect $Z'$ to be light
$\sim 1$ TeV.

If the condensates like $\langle \overline{T}_i
T_j \rangle$ do form, however, they are likely to make $Z'$ much heavier
than 1 TeV.  Denoting the strength of $\langle \overline{T}_i T_j \rangle$
by $\Lambda_c^2$, if $\Lambda_c \sim \Lambda_H$,
where $\Lambda_H$ is the confinement-scale of the hidden-sector,
we would expect $\Lambda_H$ and thus $Z'$ to be either
superheavy
$\sim 10^{15}$-$10^{16}$ GeV, or at least medium-heavy $\sim 10^{8}$-$10^{13}$
GeV (see below).

The mass of the $Z'$-boson is correlated with the
proton decay-rate. The heavier the $Z'$, the faster is the
proton-decay. Looking at Table 4, we see that the strength
of the effective $d=4$ operators ($\Ubar \,\Dbar \,\Dbar $ etc.) is
given by $\left(\langle
\widetilde{\overline{N_R}}/M\rangle \right)\left(\langle T_i\overline{T}_j
\rangle /M^2\right)^2 \sim 10^{-2.5}(\Lambda_c/M)^4$, and
that of the $d=5$ operator ($QQQL/M$) is given by
$\left(\langle T_i\overline{T}_j
\rangle /M^2\right)^2 \sim (\Lambda_c/M)^4$.
The observed bound on the former ($\eta_{1,2} \leq 10^{-12}$)
implies a rough upper limit of $(\Lambda_c/M)^4\leq 10^{-9.5}$
and thus $\Lambda_c \leq 10^{15.5}$ GeV, while that on the
latter (i.e. $\lambda_{1,2} \leq 10^{-7}$) implies that
$\Lambda_c \leq 10^{16.2}$ GeV, where, for concreteness,
we have set $M = 10^{18}$
GeV.

Thus, if $\Lambda_c  \leq 1$ TeV, $Z'$ would be light $\sim$
1 TeV, and accessible to LHC and perhaps NLC.
But, for this case, or even if $\Lambda_c$ is as heavy as $10^{15}$ GeV (say),
proton-decay would be too slow ($\tau_p \geq 10^{42}$ yrs.) to be observed.
On the other hand, if $\Lambda_c \sim 10^{15.4} - 10^{15.6}$ GeV,
the $Z'$-bosons would be inaccessible; but proton decay would be observable
with a lifetime $\sim 10^{32}$-$10^{35}$ years \cite{72}.
To see if such a superheavy $\Lambda_c$ is feasible, let us recall the
discussion in Sec. 5, where it was noted that an intermediate
unified coupling $\alpha_X \approx 0.2$ at $M_X \sim
10^{17}$ GeV (as opposed to the MSSM-value of $\alpha_X \approx 1/26$)
is desirable to stabilize the dilaton and that such a value
of $\alpha_X$ would be realized if there exists a vector-like
pair of families having the quantum numbers of $16+\overline{16}$
of $SO(10)$, in the $1-100$ TeV-region \cite{40}. With
$\alpha_X \approx 0.16$-$0.18$ (say), and a hidden sector gauge symmetry
like $SU(4)_H$ or $SU(5)_H$ \cite{64}, a confinement scale
$\Lambda_H \sim \Lambda_c \sim 10^{15.5}-10^{16}$ GeV would in fact
be expected.  Thus, while
rapid proton decay is prevented by string-derived symmetries of the
type discussed here \cite{14},
{\it observable rate for proton decay ($\tau_p \sim 10^{32}$-$10^{34}$
yrs.), which would be accessible to Superkamiokande and ICARUS,
seems perfectly feasible and natural} \cite{73}.

In summary, $\hat{Q}_{\psi}$ is
a good example of the type of symmetry that can
safegaurd, in conjunction with $B-L$ or $\hat{Q}_\chi$, proton-stability
from {\it all angles}, while permitting
neutrinos to have desired masses. It even helps eliminate the
potential danger from contribution of the color-triplets in the heavy tower
of states.  In this
sense, $\hat{Q}_{\psi}$ plays a very desirable role. I do not,
however, expect it to be the
only choice. Rather, I expect other string-solutions to exist,
which would yield symmetries like $\hat{Q}_{\psi}$, serving
the same purpose \cite{74}. At the same time, I feel that {\it
emergence of symmetries like $\hat{Q}_{\psi}$ is a very
desirable constraint that should be built into the searches for
realistic string-solutions.}

To conclude this section, the following remark is in order. For the sake
of argument, one might have considered an $SO(10)$-type
SUSY grand unification by including $126$ of Higgs to break
$B-L$ and ignoring string-theory constraints \cite{61}. One
would thereby be able to forbid the $d=4$ operators and give desired masses
to the neutrinos \cite{60}.
But, as mentioned before, the problems of finding a compelling
solution to the doublet-triplet splitting as well as to the
gravity-linked $d=5$ operators would still remain. This is
true not just for SUSY $SO(10)$, but also for SUSY $E_6$, as well
as for the recently proposed SUSY $SU(5) \times SU(5)$ -
models \cite{75}.
{\it By contrast, a string-derived non-GUT model, possessing a symmetry like
$\hat{Q}_{\psi}$, in conjunction with $B-L$ or $\hat{Q}_\chi$, meets
naturally all the constraints
discussed above}. This shows that string theory
is not only needed for a unity of all forces, but also for
ensuring {\it natural consistency} of SUSY-unification with two
low-energy observations --
proton stability and light masses for the neutrinos.\\

\section{Summary and Concluding Remarks}

Turning now to a summary of the first part of this talk,

$\bullet \;$ I noted in sections 1 and 2 that non-conservations of
baryon and lepton numbers are implied on the one hand by ideas of
higher unification \cite{1,2,3}, and on the other hand, by the need
for baryogenesis \cite{8,9} and by neutrino-masses as well.  The
latter two in fact suggest some form of violation of $B-L$, which,
very likely, includes a violation through heavy Majorana masses of
the right-handed neutrinos.

$\bullet \;$ If $\Delta(B-L) = 0$ decay modes of the nucleon (i.e. $p \to
\bar{\nu} K^+, \mu^+ K^o, \bar{\nu} \pi^+, e^+ \pi^o,$ etc.) turn out to
be the only observed source of non-conservation of $B$ and $L$, as
opposed to $\Delta(B-L) \neq 0$-transitions exhibited in Table 1 (i.e.
$p \to e^- \pi^+ \pi^+, n \leftrightarrow \bar{n}$ and $nn \to ppe^- e^-$
etc.), there would be no signal for new physics at about 100 TeV.
That would of course conform with conventional wisdom, which is
based on \underline{simple} mechanisms for symmetry-breaking of SUSY
GUTS and/or string-derived solutions, obtained to date.  On the other
hand, if the $\Delta (B-L) \neq 0-$ transitions such as $p \to e^- \pi^+
\pi^+$ or $n-\bar{n}-$ oscillation do show at some level, that would
clearly point to new physics at low intermediate scales $\sim 100 TeV$.
This will be counter to conventional thinking, but just for that reason
that may be quite revealing.  I comment on this issue further in
the following.

$\bullet \;$ Confining to the $\Delta (B-L) = 0$ decay modes of the
proton, assuming that they are discovered at SuperKamiokande and/or
ICARUS, we will learn much from knowing which decay-channels are
prominent or dominant.  \underline{First}, prominence of
$\bar{\nu}K^+$ and/or
$\mu^+ K^o$-mode would be a {\it strong evidence} in favor of the
dominance of $d=5$ over the $d=6$-operators, and thereby in favor of
supersymmetry, though not necessarily for SUSY GUTS.  Prominence of
the $\mu^+ K^o$ -mode would suggest large $tan \beta$ \cite{36b}.
\underline{Second},
prominence of $\bar{\nu}_\mu \pi^+$ and/or $\mu^+ \pi^o$-mode, together
especially with {\it non-observation} of the $e^+ \pi^o$-mode
would also provide the same signal.  \underline{Third},
prominence of the $e^+
\pi^o$-mode would favor, in the context of supersymmetry, either the
dominance of string-derived $d=4$ over $d=5$-operators (see Sec. 6 and
Ref. 14), or, for example, the flipped $SU(5) \times U(1)'$ -model
\cite{37a}.  It would, of course, be compatible also with
non-supersymmetric GUTS, whose unification-scales are raised, for
example, through enriched Higgs-systems, so that the associated
lifetimes exceed $10^{33}$ yrs.

It is worth commenting on the issue of $\Delta(B-L) =
0$ versus $\Delta(B-L) \neq 0$ -transitions.  Certain guidelines of
simplicity, noted below, seem to suggest that the latter would be too
slow to be
observed.  First, the straightforward meeting of the
three gauge couplings, obtained in the context of either MSSM or ESSM, and the
associated predictability of $sin^2 \theta_W$ (see Sec. 5) would be lost, if
one introduces low intermediate scale-physics (necessary for prominence
of $\Delta(B-L) \neq 0$ -transitions) \cite{25,26}, and thereby somewhat
arbitrary multi-stage running of the couplings.  Second, neutrino
masses that are relevant, especially for the MSW solution of the
solar neutrino-puzzle and for $\bar{\nu}_\tau$ being hot dark matter,
suggest a superheavy, rather than a low intermediate-scale, Majorana
mass for the right-handed neutrinos (see Secs. 2 and 6).  Furthermore,
in contrast to low intermediate-scales, such superheavy scales (e.g.
$<\widetilde{\overline{N_R}}> \sim 10^{15.5}$ GeV, see Sec. 6) do arise
naturally in the context of string-solutions (see e.g. Ref. 53); and,
they do not alter the simple running of three gauge
couplings, except \underline{near} the point of unification, which may be
good anyway to remove the mismatch between MSSM and string-unifications
(see Sec. 5).  Thus, if one is permitted to possess a {\it theoretical
prejudice}, based on grounds of simplicity, as narrated above, it would
seem that low intermediate-scale physics of a nature that would lead to
observable rates for $\Delta(B-L) \neq 0$-modes is not likely to be
realized, at least in the context of current level of thinking.

Nevertheless, I believe that it is essential to keep an open mind as
regards the planning of experiments, precisely to find out if one's
prejudices are true after all.  And, what if these prejudices turn out
to be wrong?  That has happened in the past.  A case to the point is CP-
violation.  Imagine that CP-violation was not discovered in 1964 and
that one did not know that it is needed to implement baryogenesis.
As late as the
early 70's, judged purely from the point of view of theoretical models,
based on just (u,d,s,c)-quarks, there was no compelling {\it
theoretical}
motivation for CP-violation.  In the context of renormalizable gauge
theories, one would have had to introduce extra Higgs-scalars, new gauge
interactions, complex Yukawa couplings and/or a third family
of quarks and leptons to implement CP-violation.  Apriori, it would
have appeared to be an unnecessary complication to introduce such extra
matter and/or extra parameters.  Judged from this point of view,
CP-violation might have appeared unnatural or unlikely.  Yet,
CP-violation was discovered; so was its need to implement
baryogenesis; and also
the third family.  It is, however, revealing to note in this context
that even now we do not know the precise origin of CP-violation.  There
is some analogy of this case with that of $(B-L)$-violation.  We do know
from baryogenesis and neutrino masses that, very likely, $B-L$ is
violated (see Sec. 2).  The issue that needs to be settled,
notwithstanding the question of naturalness and simplicity of present
theoretical models, is whether the rates for $(B-L)$-violating
transitions would lie in an observable range.  {\it There is no other
way to settle this question except to search for such transitions with
the highest possible precision.}

I, therefore, believe that experimental searches for both
$(B-L)$-conserving--i.e. $p \to \bar{\nu} K^+, \bar{\nu} \pi^+, \mu^+
K^o, e^+ \pi^o$ etc. -- as well as $(B-L)$-violating processes--i.e. $p
\to e^- \pi^+ \pi^+, n \leftrightarrow \bar{n}$-oscillation,
neutrinoless double beta decay, etc.--are strongly motivated.  For this
reason, I rejoice in the starting of the SuperKamiokande and look
forward to the starting of ICARUS.  I also greatly welcome proposal to
set up searches for $n-\bar{n}$ oscillation here at OakRidge, which aim
to probe into oscillation-periods of $10^{10} - 10^{11}$ sec.

The main purpose of this talk has been to address two issues:

(i)  How to resolve the mismatch between MSSM and
string-unification?, and especially,

(ii) How to prevent, {\it naturally}, rapid proton-decay in
supersymmetry?

With regard to the first, I noted some alternative possibilities.  Among
these, the only one that has a chance of being directly tested at future
high energy accelerators, including the LHC and the NLC, especially if
the two vector-like families have masses below about 1.5 TeV, is the
extended supersymmetric standard model (ESSM), which proposes a
{\it semi-perturbative unification} with intermediate unified
coupling $(\alpha_X \sim .2)$ in four dimensions \cite{40}.

With regard to the second issue, I have stressed that

$\bullet \;$ the extreme smallness of the strengths of
the $d=4$ (i.e. $\eta_i \leq 10^{-12}$) and the color-triplet mediated
and/or gravity-linked $d=5$ operators (i.e. $\lambda_{1,2} (M_{st}/M) <
10^{-7}$) deserves a natural explanation.  The problem in this regard is
somewhat analogous to that of understanding the smallness (or the
vanishing) of the cosmological constant.  Rather than being merely
accommodated by a {\it choice} of the Higgs multiplets and discrete
symmetries, the small parameters associated with the $d=4$ and the $d=5$
operators should emerge as a
{\it compelling feature}, owing to {\it symmetries} of the
underlying theory, which would provide the desired protection against
these unsafe operators.

$\bullet \;$  The symmetries in conventional SUSY GUTS
including $SO(10), E_6$ and $SU(5) \times SU(5)$ do not, however,
suffice for the purpose--especially in the matter of suppressing
naturally the $d=5$ proton-decay operators.  By contrast, I showed
that a certain {\it string-derived
symmetry}, which cannot arise within SUSY GUTS as mentioned above, but
which does arise within a class of {\it three-generation
string-solutions}, possessing non-GUT symmetries,
suffices, in conjunction with $B-L$, to safeguard
proton-stability from all potential dangers, including those which may
arise from higher dimensional operators and the color-triplets in the
heavy infinite tower of states \cite{14}.  At the same time, the symmetry in
question permits neutrinos to acquire appropriate masses.  We thus
see that {\it just seeking for an understanding--as opposed to
accommodating--proton-stability, in the context of supersymmetric
unification, drives us to the conclusion that, at the fundamental level, the
elementary particles must be string-like and not
point-like} \cite{fundamental}.  It seems
remarkable that just a low-energy observation that proton is so stable,
together with the demand of naturalness in understanding this feature,
can provide us with such a deep insight.

$\bullet \;$ It is intriguing that one needs a {\it family-dependent symmetry},
like $\hat{Q}_\psi$, to achieve the desired protection against rapid
proton decay, which cannot be realized for one- or two-family
string-solutions, but which does emerge for a class of solutions
possessing three families.

$\bullet \;$  A related remark:  the necessity of such a
family-dependent symmetry, which cannot arise {\it within} conventional
GUT symmetries, as well as the lack of a compelling mechanism for
doublet-triplet splitting in SUSY GUT theories (string-derived or not)
suggest that {\it the flavor-color gauge symmetry below the string-scale
is very likely a non-GUT string-derived symmetry like} ${\cal G}_{2311}$
{\it or} ${\cal G}_{224}$ , {\it or even flipped
$SU(5) \times U(1)'$, rather
than a GUT-symmetry like $SU(5)$, or $SO(10)$ or $E_6$.}  Recall
that, owing to string-constraints \cite{39}, the benefits of coupling
unification, quark-lepton unification and quantization of electric
charge still hold for the former, just as they do for the latter.

$\bullet \;$  Last but not least, as discussed in Sec. 6,
it is interesting that, while
symmetries like $\hat{Q}_\psi$ provide the desired protection against
rapid proton decay,
observable rates for proton-decay (i.e. lifetimes of order
$10^{32}-10^{34}$ yrs) are nevertheless perfectly natural in the context
of these solutions, provided, however, one assumes (at least for the
class of solutions considered in Ref. \cite{14}), a
{\it semi-perturbative unification}
with intermediate unified coupling $(\alpha_X \sim .18-.2)$ \cite{40}.
As discussed in Sec. 5, such intermediate coupling is suggested by ESSM, and it
may well be needed to help stabilize the dilaton, while
retaining the benefits of coupling-unification.

To conclude, the original motivations for a unity of the fundamental
forces and that for questioning baryon and lepton-number
conservations \cite{1,2,3} still persist.  Supersymmetry and
superstrings, while retaining these motivations, provide a new perspective
with regard to both issues.  As discussed here, several models of SUSY
GUTS and superstring-derived models do in fact suggest that proton decay
should occur at a rate that is accessible to ongoing searches.
Observation of proton decay would strengthen our belief in an underlying
unity of all matter and of its forces.  Determination of its dominant
decay modes, would provide us with a {\it wealth of knowledge}
regarding new physics
at very short distances, spanning from $10^{-19}$ to $10^{-32}$ cm.  The
question now is an experimental one:  Will proton decay be discovered at
SuperKamiokande and/or ICARUS?

{\bf Acknowledgements}

I wish to thank M. Dine, P. Langacker, R.N. Mohapatra,
N. Polonsky, A. Raisin, F. Wilczek, E. Witten and
especially K.S. Babu, K. Dienes, and A. Faraggi for most
helpful discussions and communications on topics included
in this talk.  I wish to thank
the organizers, especially Y. Kamyshkov and R.N. Mohapatra, for arranging a
very fruitful meeting and for their hospitality.  I also wish to thank
Delores Kight for her care and efforts in typing this manuscript.




\begin{table}
\centering
\begin{tabular}{|c|c|c|c|c||c|c|c||c|c|}
\hline
Family & States & $Q_1$ & $Q_2$ & $Q_3$ & $Q_4$ & $Q_5$ & $Q_6$ &
$\hat{Q}_{\psi}$ & $\hat{Q}_{\chi}$ \\ \hline
 & $q_1$ & $1/2$ & $0$ & $0$ & -$1/2$ & $0$ & $0$ & $1/2$
 & -$1/2$ \\
$1$ & $L_1$ & $1/2$ & $0$ & $0$ & $1/2$ & $0$ & $0$ & $1/2$
& $3/2$ \\
 & $(\Ubar , \overline{E})_1$ & $1/2$ & $0$ & $0$ & $1/2$ & $0$ & $0$ &
 $1/2$ & $3/2$ \\
  & $(\Dbar , \overline{\nu_R})_1$ & $1/2$ & $0$ & $0$ & -$1/2$ & $0$ & $0$ &
 $1/2$ & -$1/2$ \\  \hline
  & $q_2$ & $0$ & $1/2$ & $0$ & $0$ & -$1/2$ & $0$ & $1/2$
  & -$1/2$ \\
$2$ & $L_2$ & $0$ & $1/2$ & $0$ & $0$ & $1/2$ & $0$ & $1/2$
  & $3/2$ \\
 & $(\Ubar , \overline{E})_2$ & $0$ & $1/2$ & $0$ & $0$ & $1/2$ & $0$ & $1/2$
  & $3/2$ \\
 & $(\Dbar , \overline{\nu_R})_2$ & $0$ & $1/2$ & $0$ & $0$ & -$1/2$ & $0$ & $1/2$
  & -$1/2$ \\ \hline
 & $q_3$ & $0$ & $0$ & $1/2$ & $0$ & $0$ & -$1/2$ & -$1$ &
 -$1/2$ \\
$3$ & $L_3$ & $0$ & $0$ & $1/2$ & $0$ & $0$ & $1/2$ & -$1$ &
 $3/2$ \\
 & $(\Ubar , \overline{E})_3$ & $0$ & $0$ & $1/2$ & $0$ & $0$ & $1/2$ & -$1$ &
 $3/2$ \\
 & $(\Dbar , \overline{\nu_R})_3$ & $0$ & $0$ & $1/2$ & $0$ & $0$ & -$1/2$ & -$1$ &
 -$1/2$ \\   \hline \hline
 & & & & & & & & & \\
Color & $D_{45} = ({\bf 3},-2/3, 1_L, 0)$ & -$1/2$ &
-$1/2$ & $0$ & $0$ & $0$ & $0$ & -$1$ & -$1$ \\
Triplets & $\overline{D_{45}} = ( 3^{\ast},+2/3, 1_L, 0)$ & $1/2$ &
$1/2$ & $0$ & $0$ & $0$ & $0$ & $+1$ & $+1$ \\
 & & & & & & & & & \\ \hline
 & & & & & & & & & \\
 & $\overline{h_1} = ({\bf 1},0,{\bf 2_L},1/2)$ & -$1$ &
$0$ & $0$ & $0$ & $0$ & $0$ & -$1$ & -$1$ \\
Higgs & $\overline{h_2} = ({\bf 1},0,{\bf 2_L},1/2)$ & $0$ &
-$1$ & $0$ & $0$ & $0$ & $0$ & -$1$ & -$1$ \\
doublets & $\overline{h_3} = ({\bf 1},0,{\bf 2_L},1/2)$ & $0$ &
$0$ & -$1$ & $0$ & $0$ & $0$ & $+2$ & -$1$ \\
 & $\overline{h_{45}} = ({\bf 1},0,{\bf 2_L},1/2)$ & $1/2$ &
$1/2$ & $0$ & $0$ & $0$ & $0$ & $1$ & $1$ \\
 & & & & & & & & & \\ \hline
 & & & & & & & & & \\
 & $V_1, \overline{V}_1$ & $0$ & $1/2$ & $1/2$ & $1/2$ & $0$ & $0$ &
 -$1/2$ & $2$ \\
 & $T_1, \overline{T}_1$ & $0$ & $1/2$ & $1/2$ & -$1/2$ & $0$ & $0$ &
 -$1/2$ & $0$ \\ \cline{2-10}
 & & & & & & & & & \\
Hidden & $V_2, \overline{V}_2$ & $1/2$ & $0$ & $1/2$ & $0$ & $1/2$ & $0$ &
 -$1/2$ & $2$ \\
Matter & $T_2, \overline{T}_2$ & $1/2$ & $0$ & $1/2$ & $0$ & -$1/2$ & $0$ &
 -$1/2$ & $0$ \\ \cline{2-10}
 & & & & & & & & & \\
 & $V_3, \overline{V}_3$ & $1/2$ & $1/2$ & $0$ & $0$ & $0$ & $1/2$ &
 $1$ & $2$ \\
 & $T_3, \overline{T}_3$ & $1/2$ & $1/2$ & $0$ & $0$ & $0$ & -$1/2$ &
 $1$ & $0$ \\ \hline
\end{tabular}
\caption{Partial List of Massless States from Ref. [69].
(i) The quark and lepton fields have the standard properties
under $SU(3)^C \times U(1)_{B-L} \times SU(2)_L \times U(1)_{I_{3R}}$,
which are not shown, but those of color triplets and Higgses are
shown. (ii) Here $\hat{Q}_{\psi} \equiv Q_1 + Q_2 -2Q_3$
and $\hat{Q}_{\chi} = (Q_1 + Q_2 + Q_3) + 2(Q_4 + Q_5 + Q_6)$ (see Eq.
(14)).  (iii) The doublets $\overline{h}_{1,2,3,45}$ are
accompanied by four doublets $h_{1,2,3,45}$ with quantum numbers
of conjugate representations, which are not shown. (iv) The
$SO(10)$-singlets $ \{ \phi \} $ which possess $U(1)_i$-charges,
and the fractionally charged states which
become superheavy, or get confined, are not shown.
In Ref. [69], since only $\overline{h_1}$ and $h_{45}$ remain
light, families 1, 2 and 3
get identified with the $\tau$, $\mu$ and $e$ - families respectively.
Hidden matter $V_i, \overline{V}_i, T_i$ and $\overline{T}_i$
are $SO(10)$-singlets
and transform as $(1,{\bf 3}), (1, \overline{{\bf 3}}), ({\bf 5}, 1)$ and
$(\overline{{\bf 5}}, 1)$, respectively, under $SU(5)_H \times SU(3)_H$.}
\end{table}
\begin{table}
\centering
\begin{tabular}{|c|c|c|c|c|c|c|c|}
\hline
Operators & Family  & $Y$ & $B-L$ & $\hat{Q}_{\psi}$ &
$\hat{Q}_{\chi}$ & $\hat{Q}_{\chi}+ \hat{Q}_{\psi}$ & If  \\
 & Combinations & & & & & & Allowed \\ \hline
 & (a) All & & & & & & \\
$\Ubar \,\Dbar \,\Dbar , QL\Dbar , LL\overline{E}$
  & except (b)
& $\surd$ & $\times$ & $\times$ & $\times$ & $\times$ &
unsafe \\ \cline{2-8}
  & (b) 3 fields from  & & & & & & \\
 & 3 different & $\surd$ & $\times$ &
  $\surd$ & $\times$ & $\times$ &
unsafe  \\
 & families & & & & & & \\ \hline
 & & & & & & & \\
$(\Ubar \,\Dbar \,\Dbar$ or $QL\Dbar)(\overline{N_R}/M)$
& All & $\surd$ & $\surd$ & $\times$ & $\surd$ & $\times$
& unsafe \\
 & & & & & & & \\
$(\Ubar \,\Dbar \,\Dbar$ or $QL\Dbar) (\overline{N_R}/M)\times$
& All & $\surd$ & $\surd$
& $\surd$ &  $\times$ & $\times$ & safe \\
$[(\overline{h_1}/M)^2$ or (``$\phi$''$/M)^n]$ & & & & & & & \\
 & & & & & & & \\
$(\Ubar \,\Dbar \,\Dbar$ or $QL\Dbar) (\overline{N_R}/M)\times$
& Some($\dagger$) & $\surd$ & $\surd$ & $\surd$ & $\surd$ & $\surd$
& safe \\
$(T_i \overline{T}_j/M^2)^2$ & & & & & & & \\ \hline
 & & & & & & & \\
$QQQL/M$ & All & $\surd$ & $\surd$ & $\times$ & $\surd$ & $\times$
& unsafe  \\
$(QQQL/M)(N_L^i/M)_{i=1,2}$ & e.g.($1,2,1,3$) & $\surd$ & $\times$
& $\surd$ & $\times$ & $\times$ & unsafe  \\
$(QQQL/M)(N_L^i/M)(\overline{N}_R^j/M)$ & All & $\surd$ & $\surd$
& $\times$ & $\surd$ & $\times$ & safe(?)  \\
 & & & & & & & \\
$(QQQL/M)(T_i\overline{T}_j/M^2)^2$ & Some($\dagger$) & $\surd$ & $\surd$
& $\surd$ & $\surd$ & $\surd$ & safe  \\
$\Ubar \,\Ubar \,\Dbar \,\overline{E}/M$ & All & $\surd$ & $\surd$
& $\times$ & $\times$ & $(\ast)$ & unsafe  \\
$LL\overline{h_i}\,\overline{h_i}/M$ & All & $\surd$ & $\times$
& $\times$ & & $(\ast)$ & safe  \\
 & & & & & & & \\ \hline
\end{tabular}
\caption{The roles of $Y$, $B-L$, $\hat{Q}_{\psi}$,
$\hat{Q}_{\chi}$ and $\hat{Q}_{\chi}+ \hat{Q}_{\psi}$
in allowing or forbidding the
relevant $(B,L)$ violating operators. Check mark ($\surd$) means
``allowed'' and cross ($\times$) means ``forbidden''. The mark
$\dagger$ signifies that the corresponding operator is allowed if
either two of the four fields are in
family (1 or 2) and two are in family 3, with $i=1$ and $j=3$; or
all four fields are in family (1 or 2) with $i=1$ and $j=2$. The mark
$(\ast)$ signifies that $(\hat{Q}_{\chi}+ \hat{Q}_{\psi})$ forbids
$\Ubar \,\Ubar \,\Dbar \,\overline{E}/M$ for all family-combinations
except when all four fields belong to family 3, and that it forbids
$LL\overline{h_i}\,\overline{h_i}$ in some family-combinations,
but not in others. In labelling the operators as safe/unsafe, we
have assumed that $\langle\widetilde{\overline{N_R^i}}\rangle\sim 10^{15.5}$
GeV, $\langle\phi/M\rangle^n \leq 10^{-9}$ and $M\sim M_{st} \sim 10^{18}$ GeV,
and that hidden sector condensate-scale $\Lambda_c \leq
10^{15.5}$ GeV (see text). Note that the pairs ($Y$, $B-L$),
$(Y$, $\hat{Q}_{\psi})$, $(Y$, $\hat{Q}_{\chi})$ and $(B-L$, $\hat{Q}_{\chi})$
do not give adequate
protection against the unsafe operators. But $\hat{Q}_{\psi}$, in
conjunction with $B-L$ or $\hat{Q}_\chi$, gives adequate protection against
all unsafe operators.
This establishes the necessity of string-derived symmetries like
$\hat{Q}_{\psi}$ (which can not emerge
from familiar GUTs including $E_6$) in ensuring proton-stability.}
\end{table}


\begin{thebibliography}{999}
\bibitem{19} If there was no L-R symmetry and no $\nu_R$, ${\nu_L}$'s can
still obtain Majorana masses through gravity-induced effects, using
higher dimensional operators such as $\nu_L \nu_L \phi \phi/M$, where
$\phi$ is the Higgs doublet, and M is a characteristic mass-scale.
With $<\phi> \sim 250 GeV$, and $M \sim
10^{18} GeV$, however, one obtains $m(\nu_L) \sim 10^{-7}eV$,
which is too small to
be relevant to the MSW solution for the solar neutrino puzzle.
\bibitem{50} In Ref. 42, a mass-scale of 1 TeV was used for the masses
of the vector-like families.  This was suggested by the SUSY-preon model
(Ref. 27).  With elementary quarks, this restriction is relaxed.  This
is because, following discussions in Refs. 27 and \cite{40}, it is easy to see
that the masses of the known fermions restrict only the
\underline{ratio} of the VEVs of $H_S$ and $H_\lambda$ (see Ref. \cite{40} for
notations), where $<H_S>$ mixes chiral and vector-like families while
$<H_\lambda>$ gives diagonal masses to the latter.  I
would in this note allow their masses to lie in the range of 1-100 TeV,
which would still lead to intermediate unified coupling, as desired in
Ref. \cite{40}.  Heavier masses of order 10-100 TeV are suggested
by recent works on gauge-mediated SUSY-breaking, where vector-like
quarks serve as messengers.
\bibitem{53} Such a see-saw pattern of mass-matrix arises naturally in
the SUSY-preon model [27].  This is abstracted for elementary
quarks \cite{40} to preserve the explanation of inter-family mass-hierarchy.
It remains to be seen whether such a pattern can be derived from a
superstring theory for elementary quarks as well.
\bibitem{63} Our discussion in this section is similar in spirit,
     though not in content, to that of Weinberg in Ref. \cite{13}.
\bibitem{67a} If the problem of doublet-triplet splitting can be solved
naturally for a string-derived GUT model, and if symmetries of a nature
to be presented below can arise in these models, so as to prevent rapid
proton decay, they would of course be just as acceptable as the
string-derived non-GUT models.
\bibitem{69} As far as we can see, Yukawa couplings of the color-triplets
     in the heavy tower need not
     all be suppressed to the same extent as those of the
     electroweak doublets.
\bibitem{70} Note $\hat{Q}_{\psi}$ assigns the same charge to all
     sixteen members of a given family. In this sense,
$\hat{Q}_\psi$ (though not $\hat{Q}_\chi$) commutes with $SO(10)$.
\bibitem{72} If $< \widetilde{\overline{N}}_R > \neq 0$,
the d=4 operators (see table 4)
lead to proton decay rate $\propto
( \eta_1 \eta_2 )^2 \propto (\Lambda_c / M )^{16}$, which varies 
extremely rapidly with
$\Lambda_c$. It is interesting that, despite this rapid variation,
proton decay is observable for a plausible 
though narrow range of values of $\Lambda_c \approx 10^{15.5}$ GeV. 
If $< \widetilde{\overline{N}}_R > = 0$, as in Ref. \cite{64} (see also
remarks in Ref. \cite{68}), only the d=5 operators with strengths
$\sim ( < T_i \overline{T}_j > / M^2 )^2 $ are relevant (see table 4).
These will lead to proton decay rate $\propto ( \Lambda_c / M )^8$,
which would be observable if $\Lambda_c \sim 10^{16}$ GeV (see text).
In contrast to the $d=5$ operators, if the $d=4$ operators ($\bar{U}
\bar{D} \bar{D}$ and $QL \bar{D}$) induced as above dominate, they can
yield (through
${\tilde{d}}_R$-exchange) strange as well as non-strange particle decay
modes of the nucleon.  Thus, through these operators, the $e^+
\pi^o$-mode can be prominent or even dominant.
\bibitem{73} In all these considerations, contribution of d=6
operators are neglected, because they are strongly suppressed if the relevant
scale ($M_X$ or $M$) exceeds $10^{16}$ GeV.
\end{thebibliography}
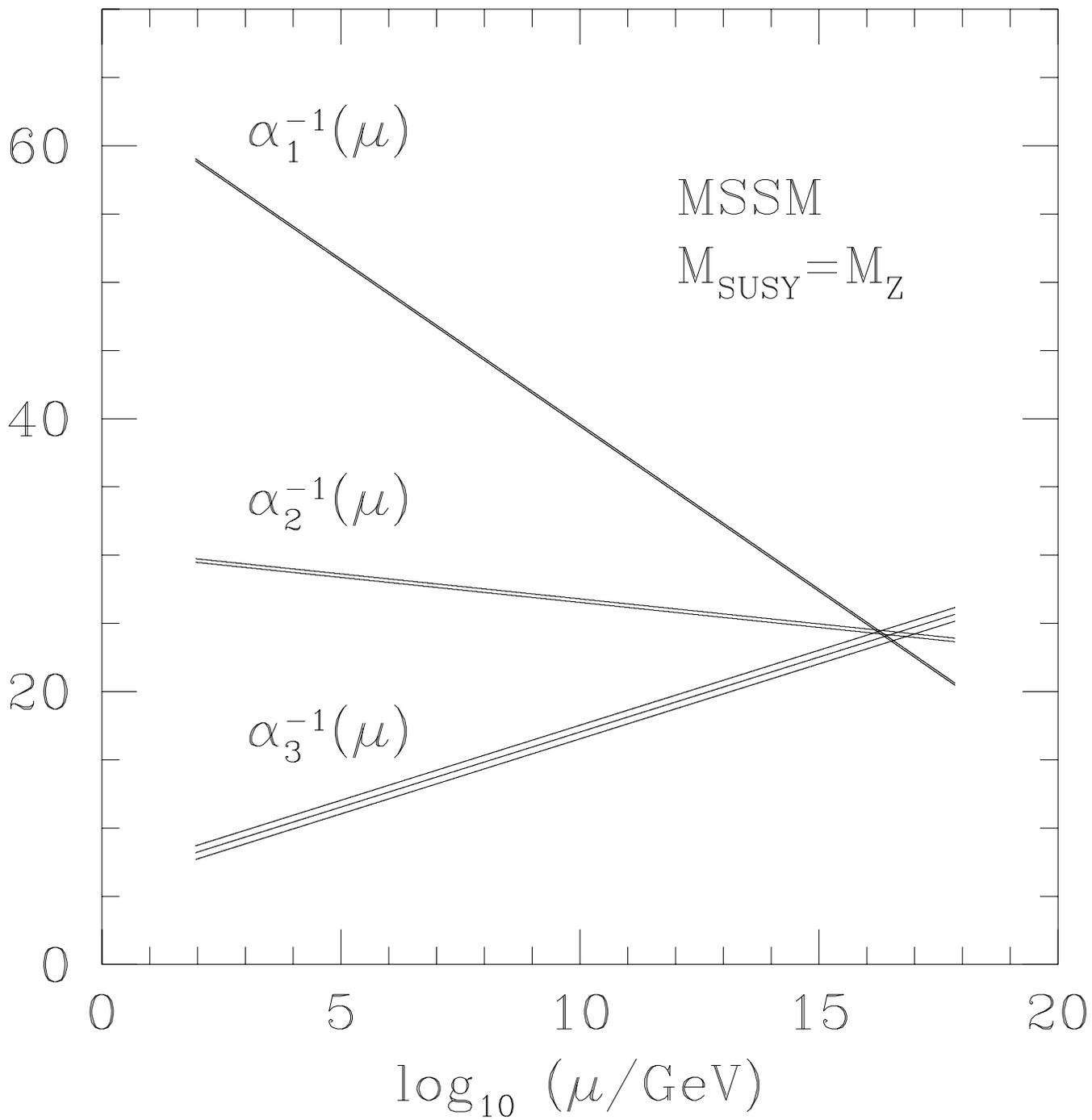
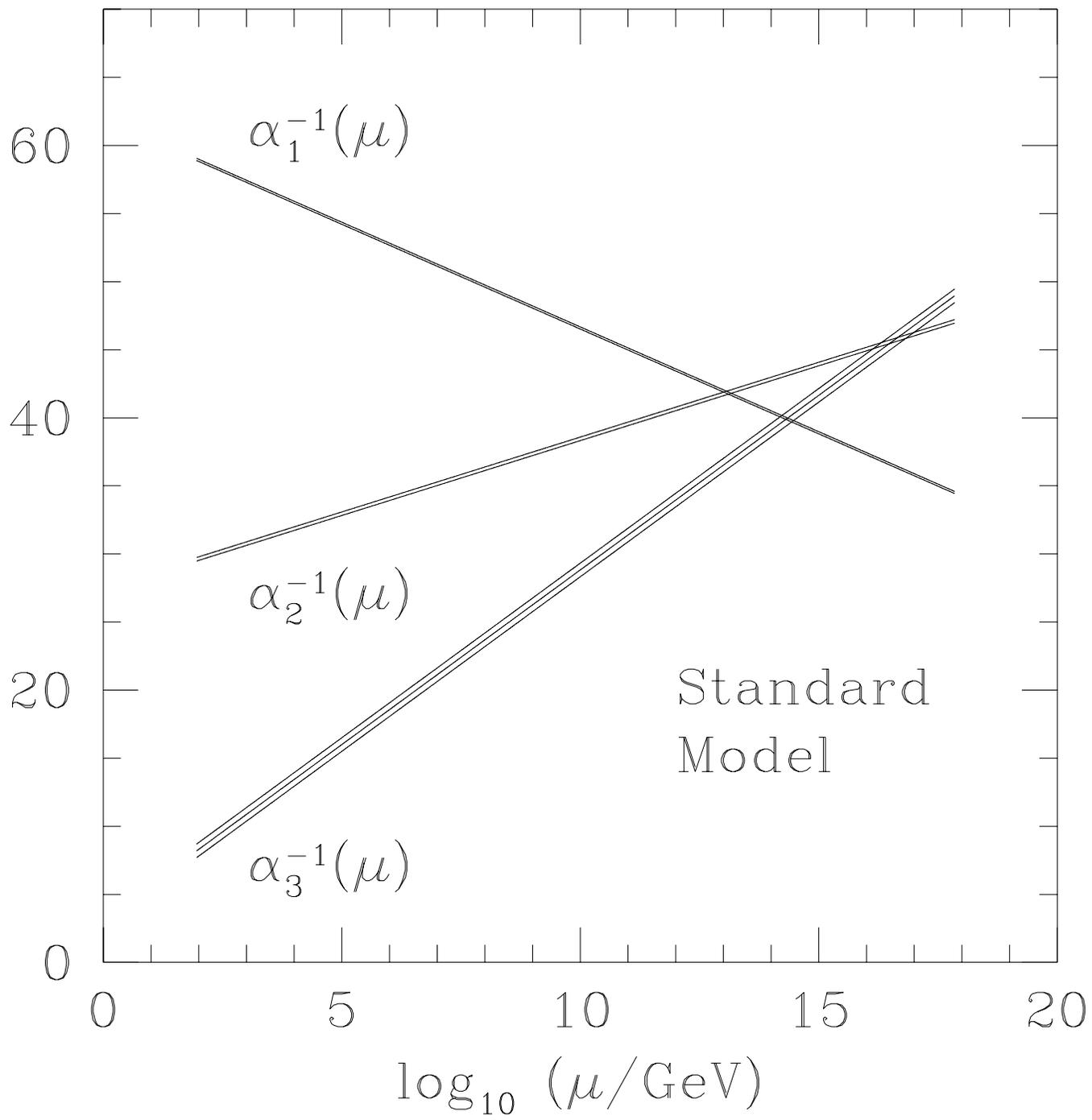
\end{document}